\documentclass[10pt,final,doublecolumn]{IEEEtran}
\IEEEoverridecommandlockouts
\usepackage{amsmath}
\usepackage{amssymb}
\usepackage{amsthm}
\usepackage{mathrsfs}
\usepackage{latexsym}
\usepackage{graphicx}
\usepackage{bbding}
\usepackage{indentfirst}
\usepackage{cases}
\usepackage{supertabular}
\usepackage{algorithm,algorithmic}
\usepackage{subeqnarray}
\usepackage{color}
\usepackage{bm}
\usepackage{stfloats}
\usepackage{subfigure}
\usepackage{array}
\allowdisplaybreaks[4]

\begin{document}
	
	\makeatletter
	\newenvironment{breakablealgorithm}
	{
			\begin{center}
				\refstepcounter{algorithm}
				\hrule height.8pt depth0pt \kern2pt
				\renewcommand{\caption}[2][\relax]{
					{\raggedright\textbf{\ALG@name~\thealgorithm} ##2\par}%
					\ifx\relax##1\relax 
					\addcontentsline{loa}{algorithm}{\protect\numberline{\thealgorithm}##2}%
					\else 
					\addcontentsline{loa}{algorithm}{\protect\numberline{\thealgorithm}##1}%
					\fi
					\kern2pt\hrule\kern2pt
				}
			}{
			\kern2pt\hrule\relax
		\end{center}
	}
	\makeatother

\title{Deep Learning-based Joint Channel Prediction and Multibeam Precoding for LEO Satellite Internet of Things }
\author{\IEEEauthorblockN{
\normalsize Ming Ying, Xiaoming Chen, Qiao Qi, and Wolfgang Gerstacker}
\thanks{Part of this paper has been presented at the IEEE ICCC 2023 [1].}
\thanks{Ming Ying and Xiaoming Chen are with the College of Information Science and Electronic Engineering, Zhejiang University, Hangzhou, 310027, China (e-mail:\{ming\_ying, chen\_xiaoming\}@zju.edu.cn). Qiao Qi is with the School of Information Science and Technology, Hangzhou Normal University, Hangzhou, 311121, China (e-mail: qiqiao1996@zju.edu.cn). Wolfgang Gerstacker is with the Institute for Digital Communications, University of Erlangen-N\"{u}rnberg, Erlangen 91058, Germany (e-mail: wolfgang.gerstacker@fau.de).}
}\maketitle

\begin{abstract}
Low earth orbit (LEO) satellite internet of things (IoT) is a promising way achieving global Internet of Everything, and thus has been widely recognized as an important component of sixth-generation (6G) wireless networks. Yet, due to high-speed movement of the LEO satellite, it is challenging to acquire timely channel state information (CSI) and design effective multibeam precoding for various IoT applications. To this end, this paper provides a deep learning (DL)-based joint channel prediction and multibeam precoding scheme under adverse environments, e.g., high Doppler shift, long propagation delay, and low satellite payload. {Specifically, this paper first designs a DL-based channel prediction scheme by using convolutional  neural networks (CNN) and long short term memory (LSTM), which predicts the CSI of current time slot according to that of previous time slots. With the predicted CSI, this paper designs a DL-based robust multibeam precoding scheme by using a channel augmentation method based on variational auto-encoder (VAE).} Finally, extensive simulation results confirm the effectiveness and robustness of the proposed scheme in LEO satellite IoT.
\end{abstract}

\providecommand{\keywords}[1]{\textbf{\textit{Index terms---}} #1}

\section{Introduction}
    The exponential growth of the number of Internet-of-Things (IoT) devices, the ever-increasing demand for practical services with stringent Quality-of-Service (QoS) requirements, and the urgent need for ubiquitous connectivity are creating difficult-to-meet challenges for the existing terrestrial communication network. Specifically, smart cities, intelligent transportation, and automatic environmental monitoring will involve billions of sensors and IoT devices which bring a huge traffic load on terrestrial communication networks. Moreover, since terrestrial networks are mainly serving for populated area, for IoT devices deployed in remote areas like oceans, mountains and deserts, terrestrial communication networks cannot provide effective wireless coverage due to the hostile environment and high construction cost.

    Recently, low-earth-orbit (LEO) satellite with low transmission latency and small propagation loss has arisen public interest by its potential as an expansion of traditional terrestrial networks to address the above issues. It can provide continuous and sufficient connectivity for districts without adequate network coverage. Driven by these advances, several companies have announced their arrangements on launching thousands of LEO satellites to space before 2030s \cite{SpaceX} \cite{Oneweb}, including SpaceX, Oneweb, Iridium, Kepler and so on. The goal of these projects is to provide seamless and high-capacity global communication services especially for areas without the coverage of terrestrial communication systems. In this context, LEO satellite IoT have been identified as an important component of sixth-generation (6G) wireless networks.

    As a key technology to provide wide range coverage of IoT devices, multibeam transmission has been widely adopted in satellite communication systems. In general, multifeed reflector antennas and phased-array antennas are two types of widely used antennas. To be more specific, geostationary-earth-orbit (GEO) satellites are mainly equipped with the multifeed reflector antennas, while phased-array antennas are more suitable for the LEO satellites due to their high flexibility and wide-angle coverage. In current satellite communication systems, multiple color reuse scheme is a commonly-used multibeam solution where adjacent beams are allocated with disjoint frequency bands (or orthogonal polarization) to mitigate the co-channel inter-beam interference \cite{color}. In such a case, the sufficient system capacity can be guaranteed by reusing the different frequency bands among isolated beams. Then, to further exploit the limited spectrum resource, the full frequency reuse (FFR) scheme has been proposed, where all beams share the same frequency band so as to improve the spectral efficiency \cite{FFR}. Unfortunately, FFR will also lead to a severe co-channel interference (CCI), especially for the devices at the edges of the beams. In this case, advanced signal processing techniques should be applied to mitigate the CCI. Specifically, CCI management can be performed at the transmitter by precoding techniques \cite{mpc}. In general, linear precoding is more compatible with LEO satellite communication systems due to lower computational complexity and good performance, while non-linear precoding schemes are hard to implement in practical systems \cite{prc}.

    To this end, downlink precoding schemes in multibeam satellite communication systems have been extensively researched \cite{prcs1}-\cite{prcs5}. In \cite{prcs1}, the researchers applied a zero-forcing (ZF) precoding scheme which was commonly used in terrestrial communication network to the multibeam satellites. A generic precoding framework was proposed in \cite{prcs2} with a class of objective functions and different power constraints for multibeam satellite systems, and the performance of the proposed scheme was close to the dirty paper coding. Based on DVB-S2X standard superframes \cite{DVB}, the authors in \cite{prcs3} designed the precoding for multicast communication in frame-based multibeam satellite. However, LEO satellites can not obtain full channel state information (CSI) for the design of multibeam precoding in practice. In this case, the authors in \cite{prcs4} have put forward two robust multibeam precoding schemes based on a typical satellite channel phase error model for two kinds of application scenarios. Furthermore, by jointly exploiting the channel features of the satellite networks and terrestrial networks, a robust multibeam precoding scheme was proposed in \cite{prcs5} for integrated satellite-terrestrial networks.

    It is worth mentioned that the aforementioned researches \cite{prcs1}-\cite{prcs5}, are all based on perfect CSIs or predetermined satellite channel models. Actually, for LEO satellite IoT, it is difficult to obtain instantaneous CSI due to the long transmission delay and large Doppler shifts. Therefore, the estimated CSIs can not be applied directly to the multibeam precoding since they are often outdated. To solve this challenge, some researchers considered to replace the timely CSIs by statistic CSIs which can be obtained by long-time observations. In \cite{prcs6}, a multiple-input multiple-output (MIMO) transmission scheme was proposed for LEO satellite communication systems, where a downlink precoder was designed to maximize the average signal-to-leakage-plus-noise ratio based on statistic CSIs. Moreover, a downlink precoding scheme of maximizing the ergodic sum rate was proposed in \cite{prcs7} by using the statistic CSIs. Yet, the error between the statistic CSIs and instantaneous CSIs may lead to the performance degradation of multibeam precoding for LEO satellite IoT, which is not what we expected. Therefore, advanced CSI acquisition methods with timeliness and precision should be explored. Fortunately, artificial intelligence (AI) provides with some considerable ideas and practical schemes to address this problem owing to its rapid development and extensive application in the field of communications. For this reason, channel prediction techniques based on AI methods have been studied recently \cite{ml1}-\cite{ml4}. Different from traditional channel estimation, channel prediction can speculate the CSIs at the next moments based on the historical CSIs, which can effectively improve the timeliness of CSIs. A machine-learning based CSI prediction method has been proposed in \cite{ml1}, where the temporal channel correlation features are extracted by the convolutional neural network (CNN) and thus the future CSI can be predicted effectively. Moreover, the researchers in \cite{ml2} have proposed a long short term memory (LSTM)-based predictor to solve the problem of channel aging in LEO satellite communication systems. Additionally, to reduce the overheads of FDD systems, a complex-valued-based DNN was designed to predict the downlink CSI directly by the uplink CSI \cite{ml3}. Further, the authors in \cite{ml4} have proposed a DL-based CSI prediction scheme for a single user terminal with CNN and LSTM in LEO satellite communication systems.

    Meanwhile, several DL-based precoding schemes have also been proposed with the development of machine learning \cite{ml4}-\cite{ml7} for the better time competitiveness. For instance, to fully exploit the spatial information, a DL-based hybrid precoding scheme for Millimeter-Wave was proposed in \cite{ml5}. Moreover, the authors in \cite{ml6} put forward a DL-based secure precoding scheme for secure transmission. Besides, a DL-based quantized phase hybrid precoder was designed in \cite{ml7}, and the simulation results demonstrated that the proposed hybrid precoder has better spectral efficiency. Eventually, for LEO satellite communication system, a supervised DL-based learning approach for hybrid precoding has been put forward in \cite{ml4}. With the proposed approach, the corresponding beam vectors can be easily generated based on the downlink CSI.

    {Unfortunately, the aforementioned DL-based precoding schemes \cite{ml4}-\cite{ml7} do not take channel prediction error into consideration. Actually, in some critical application scenarios of LEO satellite IoT, e.g., intelligent transportation and emergency communications, quality of service (QoS) in the presence of channel prediction error should be guaranteed. Therefore, an effective outage-constrained robust precoding scheme is essential to satisfy QoS requirements of LEO satellite IoT with a high probability. Yet, traditional methods for outage-constrained problems are not suitable for LEO satellite IoT for the rapid time-varying feature of channel environment. Besides, there is little work tailored to the robust multibeam precoding scheme according to predicted CSI in LEO satellite IoT. Motivated by this, we intend to design a DL-based joint channel prediction and multibeam precoding scheme with outage constraints.} The main contributions are listed in the following

    \begin{enumerate}
  	
  	 \item We propose a supervised channel prediction scheme for LEO satellite IoT, which can make full use of the downlink CSI by using the convolution layer to extract the spatial features and the LSTM layers to extract the temporal correlation features. With the proposed scheme, LEO satellites can obtain timely and accurate downlink CSI for multibeam precoding.
  	
  	 \item We also propose an unsupervised multibeam precoding scheme in LEO satellite IoT. The proposed scheme transforms the robust precoding problem under outage constraints to a deep learning problem, where the precoding output is trained to be robust against potential CSI errors.
  	
  	 \item To speculate the channel prediction error more precisely, we propose a novel channel augmentation method with variational auto-encoder (VAE) structure. More specifically, the large dataset of channel errors can be generated by VAE from a small samples of channel errors. With this method, we can address the problem more practically rather than based on the assumption of predetermined channel error model. { Simulation results show that there is around 10\% performance improvement by adopting the VAE-based method.}

   \end{enumerate}

    The remainder of this article is organized as follows. In Section II, we introduce the channel model and channel estimation method for LEO satellite IoT. Based on the system model, we propose a supervised channel prediction scheme to predict the current CSI using the CSI of previous time slots in Section III. Then, in Section IV, the outage-constrained robust precoding problem is formulated and the unsupervised
    multibeam precoding scheme with a channel augmentation
    method is presented. Next, in Section IV, we
    provide extensive simulation results to verify the effectiveness
    and robustness of the proposed two schemes. Finally, Section V concludes this paper.

    \emph{Notations}: We use bold upper letters to denote matrices and bold lower letters to denote column vectors, $(\cdot)^H$ to denote conjugate transpose, $\|\cdot\|$ to denote $L_2$-norm of a vector, $\|\cdot\|_F$ to denote Frobenius norm of a matrix, and $|\cdot|$ to denote absolute value. $J_0(\cdot)$, $J_1(\cdot)$, and $J_3(\cdot)$ are used to denote zero order Bessel function, first order Bessel function, and third order Bessel function, respectively. $\mathbb{E}\{\cdot\}$ denotes expectation, $\text{tr}(\cdot)$ denotes the trace of a matrix, ${{\mathbb{C}}^{m\times n}}$ denotes the set of $m\times n$ dimensional complex matrices, $\mathcal{N}(\mu,\sigma^2)$ and $\mathcal{CN}(\mu,\sigma^2)$ are denote the Gaussian distribution and the complex Gaussian distribution with mean $\mu$ and variance $\sigma^2$, respectively.

\section{System Model}
    \begin{figure}
	\centering
	\includegraphics [width=0.5\textwidth] {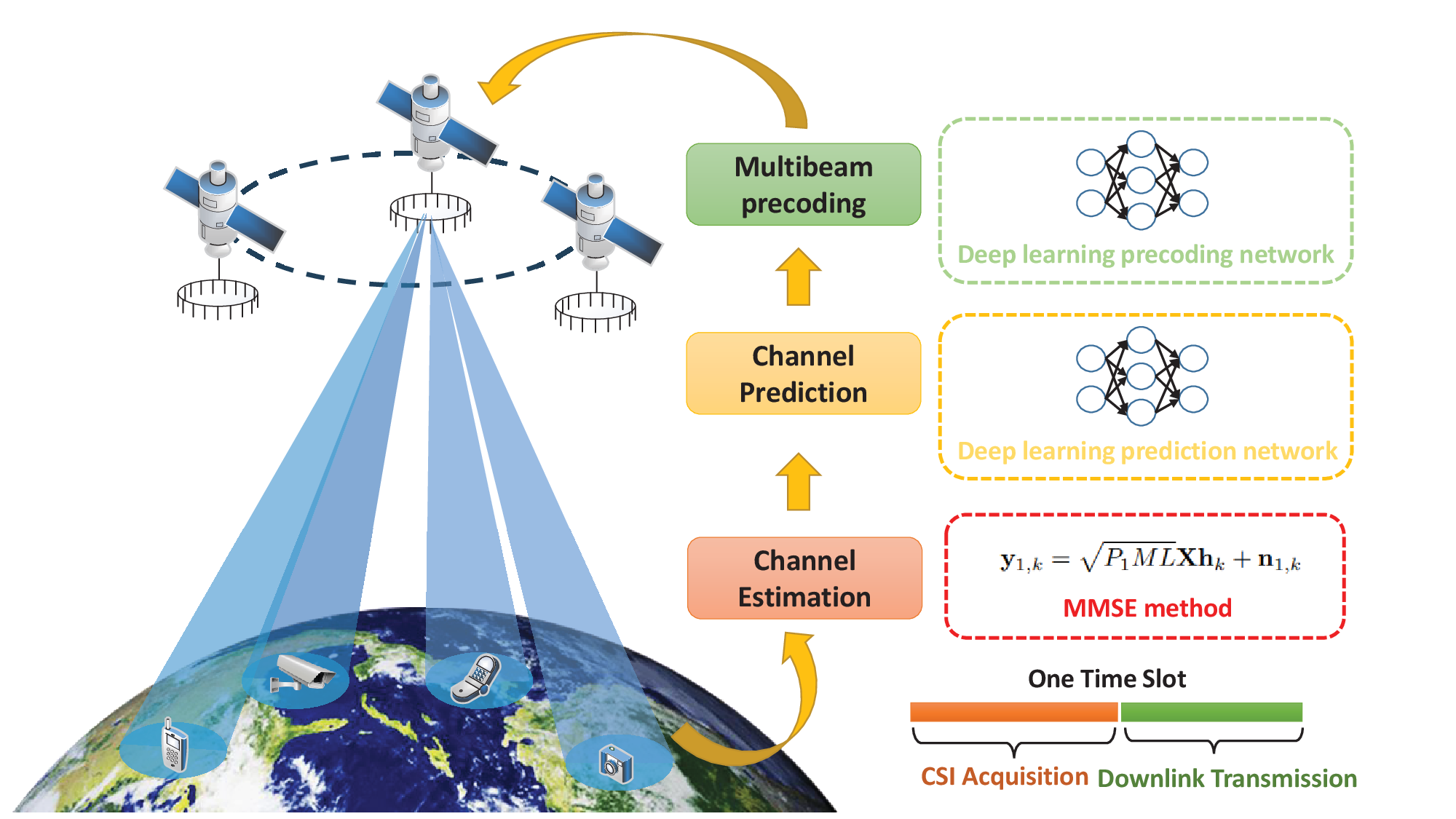}
    \caption {The proposed downlink communication architecture of LEO satellite IoT.}
    \label{Model}
    \end{figure}

    { Consider a downlink transmission scenario of LEO satellite IoT, where a satellite belonging to a LEO satellite constellation\footnote{The proposed scheme can be extended to MEO and GEO satellite constellations with appropriate adjustments.} equipped with a uniform circular array of $M$ antennas communicates $K$ widely-distributed single-antenna IoT devices\footnote{Due to low cost and small size, the IoT devices in the considered system have only one antenna. Moreover, due to limited resource and capability, the IoT devices cannot carry out complex multiuser detection on the received signal.} in its serving area over the same time-frequency resource block.} The system works in slotted time. A time slot is partitioned into two components, one for CSI acquisition, and the other for downlink transmission. As illustrated in Fig. \ref{Model}, each IoT device obtains downlink CSI via channel estimation and then conveys it to the LEO satellite through uplink channels. Due to CSI feedback delay, the LEO satellite predicts the current downlink CSI by using the estimated CSI of past time slots. With the predicted CSI, the LEO satellite generates $K$ precoded spot beams and broadcasts the precoded signals over downlink channels. In what follows, we first specify the downlink channel model, then provide a channel estimation method and the corresponding estimated CSI.

    \subsection{Channel Model}
    According to the signal propagation characteristics of the LEO satellite communications \cite{model1}-\cite{Doppler}, the downlink channel from the LEO satellite to the $k$-th device at instant $t$ over carrier frequency $f_c$ can be expressed as

    \begin{equation}\label{channel1}
	\begin{aligned}
		\mathbf{h}_k(t,f_c)=&g_k(f_c)\bigg(\sqrt{\frac{\lambda_k  }{\lambda_k+1}}\mathbf{h}^{\text{LOS}}_k(t,f_c)   \\
		&+\sqrt{\frac{1}{\lambda_k+1}}\mathbf{h}^{\text{NLOS}}_k(t,f_c)\bigg),\\
	\end{aligned}
    \end{equation}
    where $g_k(f_c)$ is the channel large-scale fading, which is given by
    \begin{equation}\label{g_k}
    g_k(f_c) =\sqrt{ (\frac{c}{4\pi f_cd_0})^2\cdot \frac{G_k\omega_k}{\kappa BT}\cdot \frac{1}{r_k}}
    \end{equation}
    where $(\frac{c}{4\pi f_cd_0})^2$ denotes the free space loss with $c$ being the light speed, $d_0$ being the propagation distance, $G_k$ being the receive antenna gain of the device $k$, $\kappa$ being the Boltzman's constant, $B$ being the carrier bandwidth, $T$ being the temperature of the received noise, and $r_k$ is the rain attenuation coefficient related to the device $k$ whose power gain in dB $r_k^{dB}=20\log_{10} r_k$, follows log-normal random distribution $\ln(r_k^{dB})\sim\mathcal{N}(\mu_r, \sigma_r^2)$. Moreover, $\omega_k$ denotes the satellite transmit antenna gain, which is given by
	\begin{equation}\label{omega_k}
		\omega_k=G_s\Big(\frac{J_1(\varphi_k)}{2\varphi_k}+36 \frac{J_3(\varphi_k)}{\varphi_k^3}\Big)^2,
	\end{equation}	
    where $G_s$ represents the maximum satellite antenna gain, and $\varphi_k = \frac{\pi d_sf_c}{c}\sin(\theta_k)$ with $d_s$ being the diameter of circular antenna array on the satellite and $\theta_k$ being the off-axis of the satellite boresight to the device $k$.

    For channel small-scale fading, $\lambda_k $ is the Rician factor, $\mathbf{h}_k^{\text{LOS}}(t,f_c)\in {\mathbb{C}}^{M\times 1}$ denotes the line-of-sight (LOS) component of LEO satellite channel, which is mainly influenced by the propagation distance $d_0$. Since the LEO satellite moves in a limited range during a time slot, it is reasonable to assume that $d_0$ is unchanged. $\mathbf{h}_k^{\text{NLOS}}(t,f_c)\in {\mathbb{C}}^{M\times 1}$ denotes the non-line-of-sight (NLOS) component of LEO satellite channel, which can be expressed as
	\begin{equation}\label{nlos}
		\begin{aligned}
	    \mathbf{h}_k^{\text{NLOS}}(t,f_c)= &\sqrt{\frac{1}{L_k}}\sum_{l=1}^{L_k}a_{k,l}\cdot \exp\{j2\pi t\nu_{k,l}\}\cdot \\
	    &\exp\{j2\pi f_c\tau_{k,l}\}\cdot\mathbf{G}_k(\varphi_k,\phi_k)\\
        \end{aligned}
    \end{equation}	
    where $L_k$ denotes the number of NLOS paths of the device $k$, $a_{k,l}$ denotes the complex channel gain of path $l$ for the device $k$,  which satisfies the equation $1/L_k\cdot\sum_{l=1}^{L_k}\mathbb{E}[|a_{k,l}|^2] = 1$, $\nu_{k,l}$ and $\tau_{k,l}$ are the Doppler shift and the propagation delay of the $l$-path of the device $k$'s channel, respectively, and $\mathbf{G}_k(\theta_k,\phi_k)$ is the array responding vector of uniform circular arrays, which is given by
   \begin{equation}\label{arv}
   	\begin{aligned}
   	\mathbf{G}_k(\varphi_k,\phi_k) = &[\exp\{j\varphi_kcos(\phi_k),\exp\{j\varphi_kcos(\phi_k-\eta_1),\\
   	&\cdots,	\exp\{j\varphi_kcos(\phi_k-\eta_{M-1})\}]\\
   \end{aligned}
    \end{equation}	
    where $\eta_m=2\pi m/M$, $\varphi_k$ is defined in (\ref{omega_k}) and $\phi_k$ is the azimuth angle of the device $k$ to the array center.

    It is widely known that the LEO satellite has fast mobility speed and long propagation distance, resulting in large Doppler shift and transmission delay.  For the Doppler shift $\nu_{k,l}$, it can be divided into two parts, i.e., $\nu_{k,l} = \nu_{k,l}^{\text{Sat}}+\nu_{k,l}^{\text{Dev}}$, where $\nu_{k,l}^{\text{Sat}}$ and $\nu_{k,l}^{\text{Dev}}$ are caused by the motion of LEO satellite and device $k$, respectively. For $\nu_{k,l}^{\text{Sat}}$, it can be seen as nearly identical for different paths due to the high altitude of the LEO satellite orbit \cite{Doppler}. Hence, we can omit subscript $l$ as $\nu_{k}^{\text{Sat}}$ for each $\nu_{k,l}^{\text{Sat}}$.
     {As the scattering characteristics around the IoT devices mainly determine the Doppler shifts caused by the mobility of IoT devices, the modeling of the Doppler shift of IoT devices in LEO satellite communications can be similar to that in the traditional terrestrial communications \cite{prcs6}. Specifically, the Doppler shifts $\nu_{k,l}^{\text{Dev}}$ are typically different for different propagation paths due to the motion of IoT devices, which contribute the Doppler spread of LEO satellite channels.} For the propagation delay $\tau_{k,l}$, we use $\tau_k^{\text{min}} = \text{min}\{\tau_{k,l}\}$ to represent the minimum value of the propagation delays for device $k$ among all the paths. Let $\tau_{k,l}^{\text{Dev}}=\tau_{k,l}-\tau_k^{\text{min}}$, then the NLOS channel component can be transformed as
    \begin{equation}\label{nlos_cp}
    	\begin{aligned}
        \mathbf{h}_k^{\text{NLOS}}(t,f_c)= &\sqrt{\frac{1}{L_k}}\sum_{l=1}^{L_k}g_{k,l}\cdot \exp\{j2\pi t\nu_{k,l}^{\text{Dev}}\}\\
        &\cdot\exp\{j2\pi f_c\tau_{k,l}^\text{Dev}\}\cdot\mathbf{G}_k(\varphi_k,\phi_k)\\
    \end{aligned}
    \end{equation}	
    In general, such a channel model can be applied to different scenarios, and relevant parameters depend on the characteristics of the specific scenario.

\subsection{Channel Estimation}
    For simplicity, we use $\mathbf{h}_k$ to denote the LEO satellite downlink CSI in a specific time slot. To gain the corresponding downlink CSI, the LEO satellite transmits a normalized pilot matrix $\mathbf{X}\in\mathbb{C}^{L\times M}$ to the IoT devices, hence the received pilot signal at the device $k$ can be expressed as
    \begin{equation}\label{downce1}
	    \mathbf{y}_{1,k}=\sqrt{P_1ML}\mathbf{X}\mathbf{h}_k+\mathbf{n}_{1,k},
    \end{equation}
    where $P_1$ is the transmit power of pilot matrix, $L$ is the length of pilot sequence, and $\mathbf{n}_{1,k}$ is the additive white Gaussian noise (AWGN) with variance $\sigma_{1,k}^2$. By applying the minimum mean square error (MMSE) channel estimation method, the device $k$ can obtain the estimated CSI $\hat{\mathbf{h}}_k$ and then conveys it to the LEO satellite. In such a case, the relationship between estimated CSI and real CSI can be expressed as
    \begin{equation}\label{es1}
 	 \begin{aligned}
 		 \mathbf{h}_k =& (\mathbf{R}_{\mathbf{h}_k}+\frac{\sigma_0^2}{SML})\mathbf{R}^{-1}_{\mathbf{h}_k}\hat{\mathbf{h}}_k+(\mathbf{X}^H\mathbf{X})^{-1}\mathbf{X}^H\mathbf{n}_k\\
 		 & =\boldsymbol{\xi}_k\cdot\hat{\mathbf{h}}_k+\mathbf{e}_{1,k}\\
 		 \end{aligned},
    \end{equation}
    where $\mathbf{R}_{\mathbf{h}_k}$ is the auto-correlation matrix of the real CSI $\mathbf{h}_k$, which can be obtained by the past CSI. $\boldsymbol{\xi}_k=(\mathbf{R}_{\mathbf{h}_k}+\frac{\sigma_0^2}{SML})\mathbf{R}^{-1}_{\mathbf{h}_k}$ is the correlation matrix between $\mathbf{h}_k$ and $\hat{\mathbf{h}}_k$, and $\mathbf{e}_{1,k}=(\mathbf{X}^H\mathbf{X})^{-1}\mathbf{X}^H\mathbf{n}\sim\mathcal{CN}(\mathbf{0},\frac{\sigma^2_0}{SML}\mathbf{I})$ is the estimation error. With the estimated CSI, the LEO satellite can carry out multibeam precoding for downlink transmission. In the following, we design a DL-based channel prediction scheme for LEO satellite IoT.

\section{DL-Based Channel Prediction}
    Due to high-speed mobility and feedback delay, it is difficult for the LEO satellite to obtain timely downlink CSI. In other words, the obtained CSI is often outdated, which may lead to performance degradation of multibeam precoding.  To this end, we advocate to perform channel prediction based on the obtained CSI of previous time slots before multibeam precoding. Since the CSI of previous time slots is available at the beginning of the current time slot, the CSI for the current time slot can be obtained timely.
    \begin{figure}
    	\centering
    	\includegraphics [width=0.5\textwidth] {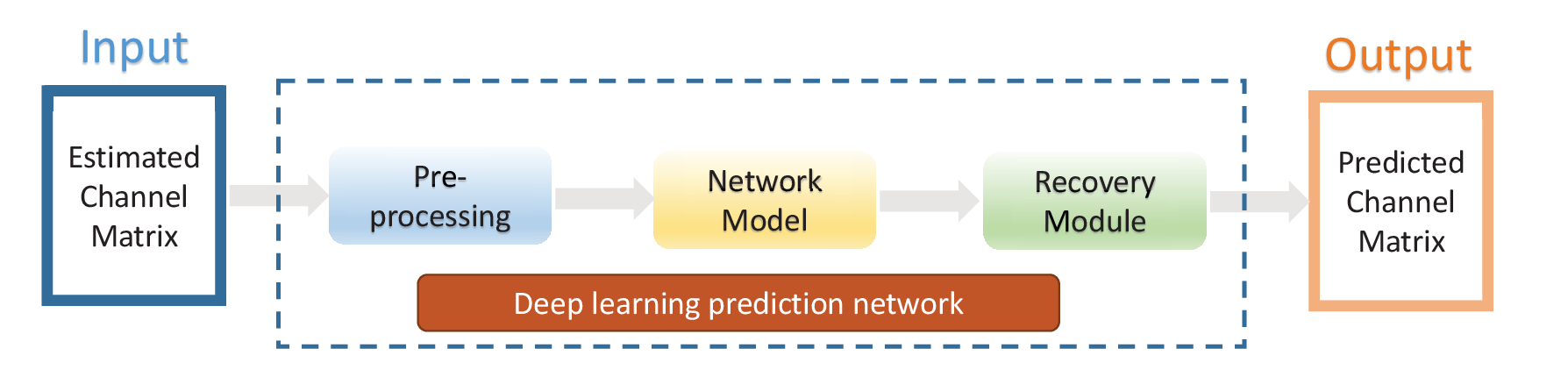}
    	\caption {The architecture of the proposed DL-based supervised channel prediction scheme.}
    	\label{SUP}
    \end{figure}
    \begin{figure}
    	\centering
    	\includegraphics [width=0.5\textwidth] {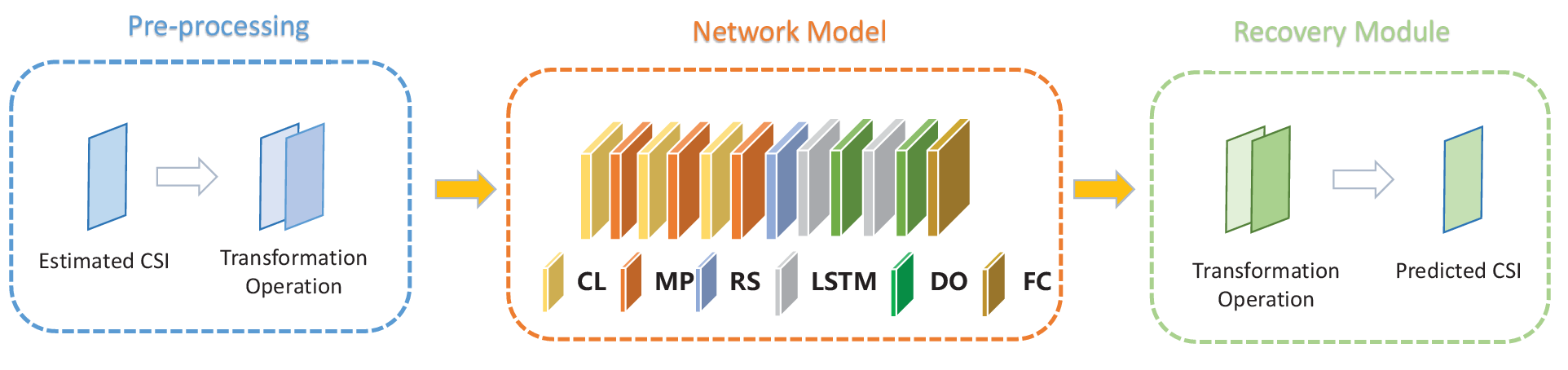}
    	\caption {The structure of deep learning prediction network.}
    	\label{network1}
    \end{figure}

    Considering the complex and time-varying characteristics of satellite-ground channels, we propose a DL-based supervised channel prediction scheme. The key of the proposed scheme is to deeply explore the temporal correlation of downlink CSI over adjacent time slots, so as to predict the current downlink CSI directly with the obtained CSI of past time slots. As illustrated in Fig. \ref{SUP}, the proposed deep learning prediction network (DLPDN) includes three main steps: 1) Input pre-processing, which aims to transform the original dataset into the form of DNN input. 2) Network model design, which extracts the significant features from the input. 3) Effective recovery, which recovers the desired information from the key features extracted from the network. In the following, we will introduce these modules in detail.

    \emph{1) Pre-processing:} For clarity, we adopt $\hat{\mathbf{H}}=[\hat{\mathbf{h}}_1, \hat{\mathbf{h}}_2,\cdots,\hat{\mathbf{h}}_k]\in\mathbb{C}^{M\times K}$ as the estimated channel matrix of all devices. In the pre-processing, to meet the required form of the DNN input, a transformation operator $\zeta(\cdot)$ is introduced to map the estimated CSI $\hat{\mathbf{H}}$ from the complex domain into the real domain, that is,
    \begin{equation}\label{realpart}
    	\zeta(\hat{\mathbf{H}})=\bigg[\begin{array}{ll}
    		\Re\{\hat{\mathbf{H}}\} \\
    		\Im\{\hat{\mathbf{H}}\} \\
    	\end{array}
    	\bigg]\in\mathbb{C}^{2M\times K},
    \end{equation}
    where $\Re\{\hat{\mathbf{H}}\}$ and $\Im\{\hat{\mathbf{H}}\}$ are the real part and imaginary part of the estimated CSI, respectively. Then, the input of DLPDN $\hat{\mathcal{H}}$ is constructed by the $\zeta(\hat{\mathbf{H}})$ with different time steps, i.e.,
    \begin{equation}\label{input}
    	\begin{aligned}
    		\hat{\mathcal{H}}=&[\zeta(\hat{\mathbf{H}}(t-\Delta t)), \zeta(\hat{\mathbf{H}}(t-2\Delta t)),\cdots,\zeta(\hat{\mathbf{H}}(t-w_{\text{step}}\Delta t))]\\
    		&\in\mathbb{C}^{w_{\text{step}}\times2M\times K} \\
    	\end{aligned}
    \end{equation}
     where $\zeta(\hat{\mathbf{H}})$ is a three-dimensional tensor, and $w_{\text{step}}$ is the number of time slots used for channel prediction, also is defined as input step size, and $\Delta t$ is the duration of a time slot.

    \emph{2) Network Model:} As shown in Fig. \ref{network1}, the architecture of the deep learning network model is composed of convolution (CL) layers, max-pooling (MP) layers, Reshape (RS) layers, LSTM layers, Dropout (DO) layers and fully-connected (FC) layers. Specifically, we adopt three CL layers with ReLU activation to extract the spatial features hidden in $\hat{\mathcal{H}}$. An MP layer is equipped after each CL layer to reduce the dimension and enhance the robustness of the extracted features. The RS layer is used to transform the high-dimensional tensor into two-dimensional (input step size and variables) matrix to satisfy the input requirement of LSTM layers. Then, two LSTM layers are added to extract the temporal correlations of the input data. To be more specific, an LSTM layer consists of a series of LSTM units
    \begin{figure}
    	\centering
    	\includegraphics [width=0.5\textwidth] {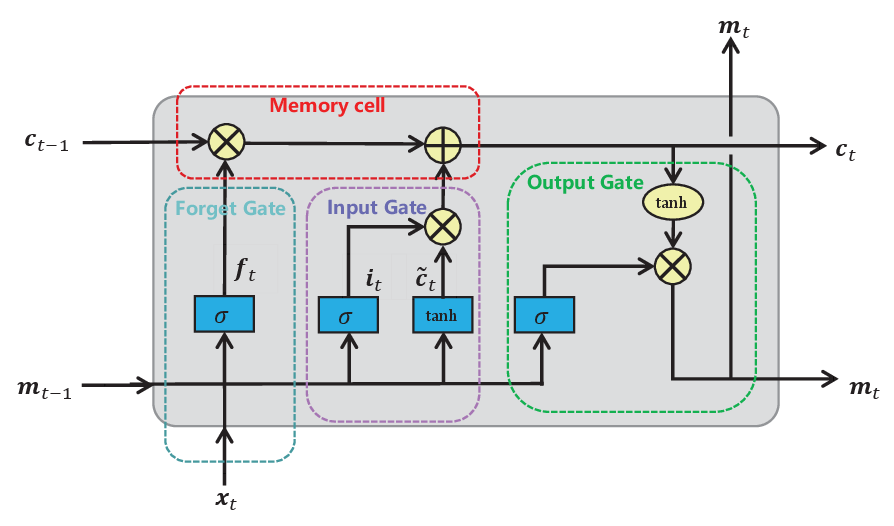}
    	\caption {The architecture of LSTM unit with different gates.}
    	\label{LSTM}
    \end{figure}
    as illustrated in Fig. \ref{LSTM}, and each unit consists of one memory cell and three gates. The forget gate determines the information kept from the previous state, and the output of the forget gate $\mathbf{f}_t$ can be expressed as
    \begin{equation}\label{LSTM1}
    	\mathbf{f}_t = \mathcal{F}_{\sigma}(\mathbf{W}_f\cdot[\mathbf{m}_{t-1};\mathbf{x}_t]+\mathbf{b}_f)
    \end{equation}
    where the $\mathcal{F}_{\sigma}$ denotes the sigmoid function, namely $\mathcal{F}_{\sigma}(x)=1/(1+e^{-x})$, $\mathbf{W}_f$ is the weight matrix, $\mathbf{b}_f$ is the bias vector, $\mathbf{m}_{t-1}$ is the output of the previous LSTM unit, and $\mathbf{x}_t$ is the input of the LSTM. Similarily, the input gate determines the new information conveyed to the cell, which are given by
    \begin{equation}\label{LSTM2}
    	\mathbf{i}_t = \mathcal{F}_\sigma(\mathbf{W}_i\cdot[\mathbf{m}_{t-1};\mathbf{x}_t]+\mathbf{b}_i),
    \end{equation}
    \begin{equation}\label{LSTM3}
    	\tilde{\mathbf{C}}_t = \mathcal{F}_\text{tanh}(\mathbf{W}_C\cdot[\mathbf{m}_{t-1};\mathbf{x}_t]+\mathbf{b}_C),
    \end{equation}
    where $\mathbf{i}_t$ contains the information which is updated by the new input vector, $\tilde{\mathbf{C}}_t$ is the candidate value of the updated cell state, and $\mathcal{F}_\text{tanh}(x)$ is the hyperbolic tangent function denoted by $\mathcal{F}_\text{tanh}(x)= \frac{e^{2x}-1}{e^{2x}+1}$. Then, based on the information from the two gates, the updated cell state $\mathbf{C}_t$ can be expressed as
    \begin{equation}\label{LSTM4}
	\mathbf{C}_t = \mathbf{f}_t\otimes\mathbf{C}_{t-1}+\mathbf{i}_t\otimes\tilde{\mathbf{C}}_t.
    \end{equation}
    Finally, the output gate outputs the effective information $\mathbf{m}_t$ according to the updated cell state $\mathbf{C}_t$ as below
    \begin{equation}\label{LSTM5}
    \mathbf{o}_t = \mathcal{F}_{\sigma}(\mathbf{W}_o\cdot[\mathbf{m}_{t-1};\mathbf{x}_t]+\mathbf{b}_o),
    \end{equation}
    \begin{equation}\label{LSTM6}
	\mathbf{m}_t = \mathbf{o}_t\otimes \text{tanh}(\mathbf{C}_t).
    \end{equation}
    where $\mathbf{o}_t$ controls the output of the new cell state $\mathbf{C}_t$.

    Due to the special structure of LSTM units, the new pivotal CSIs can be absorbed and the obsolete irrelevant CSIs can be discarded in LSTM layer, respectively. Meanwhile, we add a dropout layer to avoid overfitting by reducing the interdependence between the nodes. Finally, the last FC layer is used to transform the output to the desired vector.

    \emph{3) Recovery Module:} Finally, the DNN produces the predicted CSI, which can be defined as $\widetilde{\mathbf{h}}$. It should be noticed that the output of the DNN is a real-valued vector of the CSI, which contains the real part and the imaginary part of the predicted CSI. Therefore, we need to transform the $\widetilde{\mathbf{h}}$ into the complex domain, which is the reverse operation of $\zeta(\cdot)$, we denote as $\zeta^{-1}(\cdot)$. Then, the channel vector should be transformed into the channel matrix form by the transformation operator $\xi(\cdot)$. Therefore, the predicted CSI can be expressed as
    \begin{equation}\label{hpre}
      	\widetilde{\mathbf{H}}=\xi(\zeta^{-1}(\widetilde{\mathbf{h}}))=[\widetilde{\mathbf{h}}_1, \widetilde{\mathbf{h}}_2, \cdots, \widetilde{\mathbf{h}}_K].
    \end{equation}
    To evaluate the accuracy of the proposed DLPDN, we use  $\mathbf{e}_{2,k}$ to denote the error between estimated CSI and predicted CSI, which can be expressed as
    \begin{equation}\label{er2}
    	\mathbf{e}_{2,k}=\hat{\mathbf{h}}_k-\widetilde{\mathbf{h}}_k.
    \end{equation}
    By substituting (\ref{er2}) to (\ref{es1}), we can acquire the relationship between  real CSI and predicted CSI as
    \begin{equation}\label{er}
    	 \mathbf{h}_k =\boldsymbol{\xi}_k\cdot\widetilde{\mathbf{h}}_k+{\boldsymbol{\xi}_k\cdot\mathbf{e}_{2,k}+\mathbf{e}_{1,k}}=\boldsymbol{\xi}_k\cdot\widetilde{\mathbf{h}}_k+{\mathbf{e}_{k}},
    \end{equation}
    where $\mathbf{e}_{k}={\boldsymbol{\xi}_k\cdot\mathbf{e}_{2,k}+\mathbf{e}_{1,k}}$ is the total channel error.

    With the predicted CSI, a DL-based unsupervised multibeam precoding scheme is designed for LEO satellite IoT in the next section.

\section{DL-based Robust Scheme for Multibeam Precoding}
     In this section, we aim to provide a multibeam precoding scheme for LEO satellite IoT based on the predicted CSI. Since the predicted CSI at the LEO satellite is imperfect, it is essential to design a robust precoding scheme to guarantee the performance of downlink multibeam transmission.
\subsection{Problem Formulation}
    With the predicted CSI in (\ref{hpre}), the LEO satellite constructs the following superposition coded signal $\mathbf{x}$ and broadcasts it to the IoT devices over the downlink channels
\begin{equation}\label{downpre1}
	\mathbf{x}=\sum_{k=1}^{K}\mathbf{w}_ks_k,
\end{equation}
    where $s_k$ is a Gaussian distributed data signal of unit norm related to the device $k$, and $\mathbf{w}_k\in\mathbb{C}^{M\times 1}$ is the precoding vector for $s_k$ to enhance the desired signal and reduce the inter-device interference. Consequently, the received data signal at the device $k$ can be expressed as
\begin{equation}\label{resgnal}
	\begin{aligned}
		\mathbf{y}_{2,k}&= \mathbf{h}_k^H\mathbf{x}+\mathbf{n}_{2,k}  \\
		&=\underbrace{\mathbf{h}_k^H\mathbf{w}_ks_k}_{\text{Desired  signal}}+\underbrace{\sum_{j=1,j\neq k}^{K}\mathbf{h}_k^H\mathbf{w}_js_j}_{\text{Inter-device interference}}+\underbrace{\mathbf{n}_{2,k}}_{\text{AWGN}} \\
	\end{aligned}
\end{equation}
where $\mathbf{n}_{2,k}$ is the AWGN with variance $\sigma^2_{2,k}$. Based on (\ref{resgnal}), the signal to interference plus noise ratio (SINR) of the device $k$ can be computed as {Due to limited available radio spectrum and a large number of accessing devices, the spectrum sharing scheme is more practical compared to an FDMA scheme in LEO satellite IoT \cite{color}, \cite{prcs6}. This is because traditional FDMA schemes, which allocate specific frequency bands to individual devices, cannot support the access of a large number of IoT devices in the case of limited spectrum resource.}
\begin{equation}\label{SINR}
	\Gamma_k=\frac{\vert\mathbf{h}_k^H\mathbf{w}_k\vert^2}{\sum_{j=1,j\neq k}^{K}\vert\mathbf{h}_k^H\mathbf{w}_j\vert^2+\sigma_0^2}.
\end{equation}
    Then, the achievable transmission rate $\mathcal{I}_k$ is given by
\begin{equation}\label{rate}
	\mathcal{I}_k=\text{log}_2(1+\Gamma_k).
\end{equation}
     From (\ref{SINR}) and (\ref{rate}), it is seen that the precoding vector $\mathbf{w}_k$ has a great impact on the quality of the received signals. In order to improve the overall performance of downlink transmission for LEO satellite IoT, we intend to optimize the precoding vectors at the LEO satellite by maximize the WSR subject to QoS of each IoT devices. Considering the imperfect predicted CSI, an outage-constrained robust scheme is proposed to satisfy the QoS requirement with a high probability. Particularly, the robust multibeam precoding problem can be formulated as the following optimization problem
\begin{subequations}
	\begin{eqnarray}
	\mathcal{Q}1:	\underset{\{\mathbf{w}_k\}_{k=1}^K}{\mathop{\text{maxmize}}}\,\!\!&&\!\!\!\sum_{k=1}^{K}\alpha_k\cdot\mathcal{I}_k\label{OP1obj}\\
		\textrm{s.t.}&&\!\! \sum_{k=1}^{K}\vert\vert\mathbf{w}_k\vert\vert^2 \le P_2,\label{OP1st1}\\
		&&\!\! \text{Pr}\{\Gamma_k>\gamma_k\}\geq1-p_k^{\text{out}},\quad \forall k\label{OP1st2}
	\end{eqnarray}
\end{subequations}
    where $\alpha_k$ is the weight of the device $k$, $P_2$ represents the maximum transmit power of the LEO satellite, $\gamma_k$ is the required minimum SINR of the device $k$, and $p_k^{\text{out}}$ is the SINR outage probability threshold. It is obvious that the optimization problem (\ref{OP1obj}) is a non-convex problem, for which obtaining the global optimization solution in polynomial time seems to be impossible. First, the outage constraint (\ref{OP1st2}) is a kind of probabilistic constraints, which is difficult to be dealt with. Second, the optimization problem is not a polynomial over the optimization variables $\{\mathbf{w}_k\}_{k=1}^K$, because of which the expression is still difficult to be handled. In this context, we propose a DL-based unsupervised multibeam precoding scheme to address this challenging task.

\subsection{Robust Precoding Design With Deep Neural Network}
    \begin{figure}
    	\centering
    	\includegraphics [width=0.5\textwidth] {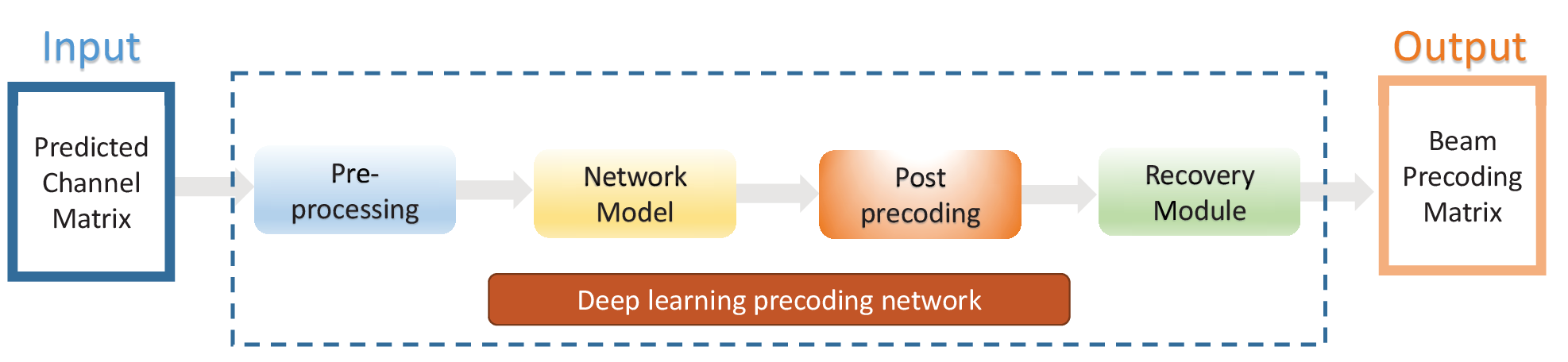}
    	\caption {The architecture of the proposed DL-based unsupervised multibeam precoding scheme.}
    	\label{framework_2}
    \end{figure}
    In this subsection, we design a DL-based unsupervised scheme to solve the formulated problem in (\ref{OP1st1}). The whole architecture is illustrated in Fig. \ref{framework_2}, where the input is the predicted channel matrix $\tilde{\mathbf{H}}$ obtained in (17), and the output is the desired precoding matrix  $\mathbf{W}=[\mathbf{w}_1,\mathbf{w}_2,\cdots,\mathbf{w}_K]$. The detailed structure of deep learning precoding network (DLPCN) will be explained later and now we focus on the problem modeled with a DL-based network. We let a DLPCN function $\mathbf{f}_{\boldsymbol{\theta}}(\cdot)$ output $\mathbf{W}$ according to the predicted CSI matrix $\widetilde{\mathbf{H}}$, which can be expressed as
    \begin{equation}\label{DNNst}
    	\mathbf{W}=\mathbf{f}_{\boldsymbol{\theta}}(\widetilde{\mathbf{H}};\boldsymbol{\theta}).
    \end{equation}
    With (\ref{DNNst}), we can rewrite some expressions in the DLPCN functional form. Specifically, the SINR in (\ref{SINR}), the weighted sum rate in (\ref{rate}), and the power constraint in (\ref{OP1st1}), can be rewritten as
    \begin{equation}\label{SINR_NN}
     	\Gamma_k(\mathbf{h}_{k},\mathbf{f}_{\boldsymbol{\theta}}(\widetilde{\mathbf{H}};\boldsymbol{\theta}))=\frac{\vert\mathbf{h}_k^H\mathbf{w}_k\vert^2}{\sum_{j=1,j\neq k}^{K}\vert\mathbf{h}_k^H\mathbf{w}_j\vert^2+\sigma_0^2},
    \end{equation}
    \begin{equation}\label{pow_NN}
 	    \mathcal{P}(\mathbf{f}_{\boldsymbol{\theta}}(\widetilde{\mathbf{H}};\boldsymbol{\theta}))=\sum_{k=1}^{K}\vert\vert\mathbf{w}_k\vert\vert^2,
    \end{equation}
  \begin{equation}\label{rate_NN}
 	\mathcal{R}(\mathbf{h}_{k},\mathbf{f}_{\boldsymbol{\theta}}(\widetilde{\mathbf{H}};\boldsymbol{\theta}))=\sum_{k=1}^{K}\alpha_k\cdot\log_2(1+\Gamma_k(\mathbf{h}_{k},\mathbf{f}_{\boldsymbol{\theta}}(\widetilde{\mathbf{H}};\boldsymbol{\theta}))).
 \end{equation}
    It is clear that the objective in (\ref{DNNst}) is to produce the solution to (\ref{OP1obj}). However, the deep learning method cannot be directly applied to solve problem $\mathcal{Q}1$, since addressing the constraints especially  the probabilistic constraint (\ref{OP1st2}) in neural network is a quite difficult task. To facilitate this problem, we introduce the concept of quantile function.

    {\emph{Definition 1:}  Given a set $X$ and a probability $\varepsilon$, for all $x\in X$, the $\varepsilon$-quantile function $Q(x;\varepsilon)$ is defined as
    \begin{equation}\label{quantile1}
    	Q(x;\varepsilon) = \text{inf}\{t|\text{Pr}(x\leq t)\leq\varepsilon\}.
    \end{equation}
    In other words, the $\varepsilon$-quantile function $Q(x;\varepsilon)$ returns a quantile value $t$ such that, for all $x$, the probability $\text{Pr}\{x\leq t\}$ is no more than the given value $\varepsilon$. With this function, given a probabilistic constraint in the following form
    \begin{equation}\label{prconst1}
	    \text{Pr}\{x\leq T\}\leq p,
    \end{equation}
    the probabilistic constraint can be rewritten in an equivalent form with an $p$-quantile function that
\begin{equation}\label{quantile2}
	T-Q(x;p)\leq 0.
\end{equation}
Applying such a property, we successfully transform the intractable probabilistic constraint into a comparison between the quantile value $Q(x;p)$ against the threshold $T$. Therefore, for the probabilistic constraint in (\ref{OP1st2}), it can be firstly rewritten as the following equivalent form
\begin{equation}\label{prcons1}
	 \text{Pr}\{\Gamma_k(\mathbf{h}_{k},\mathbf{f}_{\boldsymbol{\theta}}(\widetilde{\mathbf{H}};\boldsymbol{\theta})))\leq\gamma_k\}\leq p_k^{\text{out}},
\end{equation}
and then the constraint can be transformed as
\begin{equation}\label{prcons1}
	\gamma_k-Q(\Gamma_k(\mathbf{h}_{k},\mathbf{f}_{\boldsymbol{\theta}}(\widetilde{\mathbf{H}};\boldsymbol{\theta}));p_k^{\text{out}})\leq 0.
\end{equation}
Based on the above analysis, the constraint transformation from (24c) to (33) and (34c) is equivalent.} Hence, the problem can be further rewritten in a standard form for DNN training procedure as
\begin{subequations}
	\begin{eqnarray}
		\mathcal{Q}2:	\underset{\boldsymbol{\theta}}{\mathop{\text{minimize}}}\quad-\mathbb{E}_{\substack{\widetilde{\mathbf{H}}\in\mathbb{S}_{\widetilde{\mathbf{H}}}}}\Big[\mathcal{R}(\mathbf{h}_{k},\mathbf{f}_{\boldsymbol{\theta}}(\widetilde{\mathbf{H}};\boldsymbol{\theta}))\Big]\label{OP2obj}\\
		\textrm{s.t.}\quad\quad \mathbb{E}_{\substack{\widetilde{\mathbf{H}}\in\mathbb{S}_{\widetilde{\mathbf{H}}}}}[\mathcal{P}(\mathbf{f}_{\boldsymbol{\theta}}(\widetilde{\mathbf{H}};\boldsymbol{\theta})) ]\le P_2,\label{OP2st1}\\
		 \mathbb{E}_{\substack{\widetilde{\mathbf{H}}\in\mathbb{S}_{\widetilde{\mathbf{H}}}}}\Big[\gamma_k-Q(\Gamma_k(\mathbf{h}_{k},\mathbf{f}_{\boldsymbol{\theta}}(\widetilde{\mathbf{H}};\boldsymbol{\theta}));p_k^{\text{out}})\Big]\leq 0,\quad \forall k\label{OP2st2}
	\end{eqnarray}
\end{subequations}
where $\mathbb{S}_{\widetilde{\mathbf{H}}}$ is the set of predicted CSI matrices $\widetilde{\mathbf{H}}$. {Actually, the mathematical expectations in Eq. (34) are operated with respect to $\mathbf{\widetilde{H}}$, which is the training sample in the training procedure. It aims to make the trained DNN parameter set $\boldsymbol{\theta}$ generally applicable to all possible training samples $\mathbf{\widetilde{H}}$ in the predicted CSI dataset $\mathbb{S}_{\mathbf{\widetilde{H}}}$}. Up to now, the original optimization problem can be addressed by a DNN network. However, the deep learning method cannot output precoding vectors which automatically satisfy the constraints (\ref{OP2st1}) and (\ref{OP2st2}). For constraints which only involve the outputs of the DLPCN as (\ref{OP2st1}), we can construct some special neural network structures like Lambda-layer to satisfy these constraints. Thereby, our remaining task is to deal with the SINR constraint in (\ref{OP2st2}). Due to the complicated expression, it cannot be solved by the neural network structure. To handle this problem, we combine this constraint with the objective function together during the training procedure.  A typical method achieving this purpose is the penalty function method which uses sufficiently large positive factors to penalize the constraints. When the optimized variables violate the constraints, there will be a punishment for the objective. To be more specific, we use $\mu_k$ to penalize the constraint (\ref{OP2st2}), and the loss function of our proposed DNN can be derived as below

\begin{equation}\label{OP3obj}
	\begin{aligned}
		L(\boldsymbol{\theta})=&-\mathbb{E}_{\substack{\widetilde{\mathbf{H}}\in\mathbb{S}_{\widetilde{\mathbf{H}}}}}\Big[\mathcal{R}(\mathbf{h}_{k},\mathbf{f}_{\boldsymbol{\theta}}(\widetilde{\mathbf{H}};\boldsymbol{\theta}))\Big]  \\
		&+\sum_{k=1}^{K}\mu_k\Phi\Big(\mathbb{E}_{\substack{\widetilde{\mathbf{H}}\in\mathbb{S}_{\widetilde{\mathbf{H}}}}}\Big[\gamma_k-Q(\Gamma_k(\mathbf{h}_{k},\mathbf{f}_{\boldsymbol{\theta}}(\widetilde{\mathbf{H}};\boldsymbol{\theta}));p_k^{\text{out}})\Big]\Big) ,\\
	\end{aligned}
\end{equation}
    where $\Phi(x)$ is the function $\Phi(x)=\text{max}\{x,0\}$. Subsequently, we can transform the original optimization problem into an unconstrained problem as follows
\begin{equation}\label{Q3}
    \mathcal{Q}3:	\underset{\boldsymbol{\theta}}{\mathop{\text{minimize}}}\quad L(\boldsymbol{\theta}).
\end{equation}
	In this way, by taking the $ L(\boldsymbol{\theta})$ as the loss function for DNN,  the unconstrained problem in (\ref{Q3}) can provide an unsupervised learning approach to address the original constrained problem in (\ref{OP1obj}). Therefore, there is no need to train the model by label data.
	
\subsection{Channel Augmentation with VAE}
	We notice the output $\mathbf{W}$ of DLPCN defined in (\ref{DNNst}), only depends on the predicted CSI matrix $\widetilde{\mathbf{H}}$, but not the real CSIs $\{\mathbf{h}_k\}_{k=1}^{K}$. Yet, the DLPCN cannot address the objective term $\mathcal{R}(\mathbf{h}_{k},\mathbf{f}_{\boldsymbol{\theta}}(\widetilde{\mathbf{H}};\boldsymbol{\theta}))$ and the SINR constraint term $\Gamma_k(\mathbf{h}_{k},\mathbf{f}_{\boldsymbol{\theta}}(\widetilde{\mathbf{H}};\boldsymbol{\theta}))$ in the loss function, since the DLPCN has no real CSIs $\{\mathbf{h}_k\}_{k=1}^K$. Fortunately, the real CSIs are only involved in the evaluation of the outputs of the DLPCN, which motivates us to propose a channel augmentation method for DLPCN. During the training procedure, each predicted CSI $\widetilde{\mathbf{h}}_k$ in the predicted CSI matrix $\widetilde{\mathbf{H}}$ is augmented by a channel error $\mathbf{e}_k$ belonging to the channel error set $\mathbb{S}_{\mathbf{e}_k}$ to form a set of possible real CSI $\mathbb{S}_{\mathbf{h}_k}=\{\mathbf{h}_k|\mathbf{h}_k=\widetilde{\mathbf{h}}_k+\mathbf{e}_k,\forall \mathbf{e}_k\in\mathbb{S}_{\mathbf{e}_k}\}$. Then this channel-augmented set is used to calculate the SINR of (\ref{SINR_NN}) and the weighted sum rate of (\ref{rate_NN}), respectively.	
	\begin{figure}
		\centering
		\includegraphics [width=0.5\textwidth] {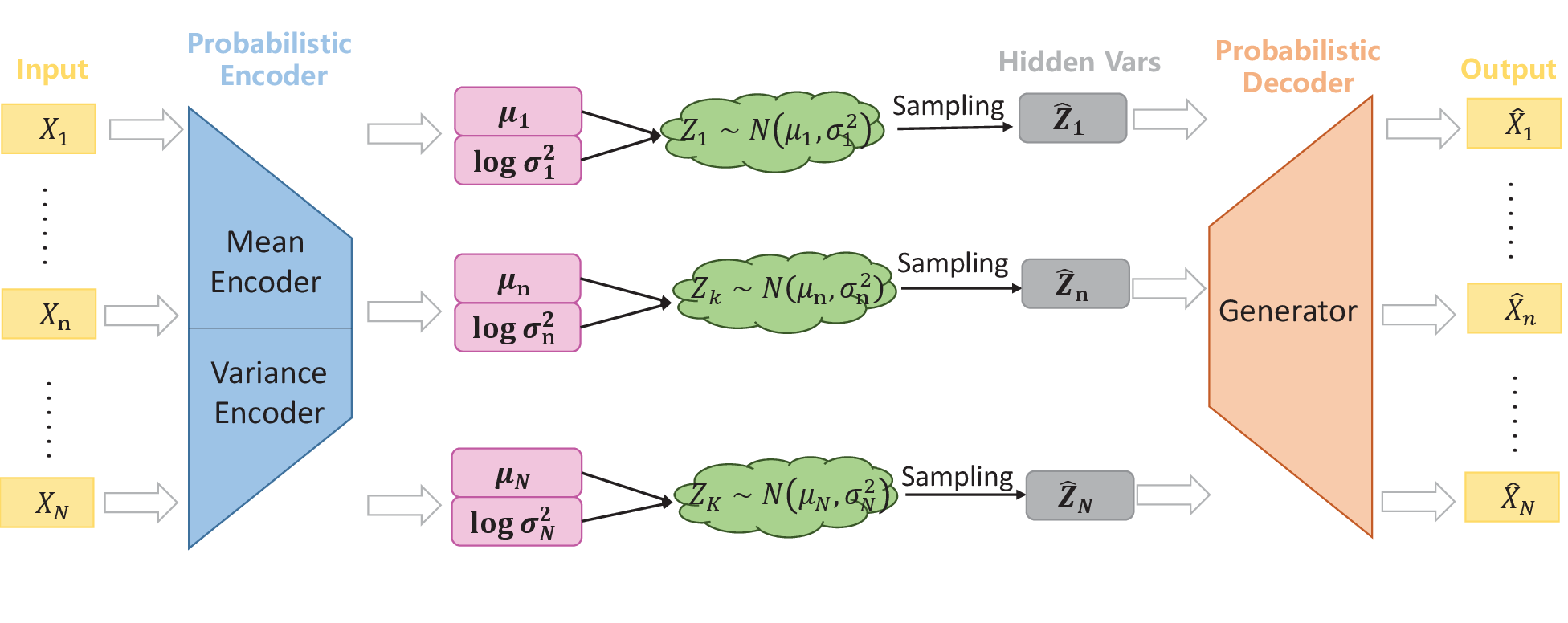}
		\caption {The structure of VAE.}
		\label{VAE}
	\end{figure}

	As mentioned in (\ref{er}), the channel error $\mathbf{e}_k$ is consisted of two parts, i.e., estimated error $\mathbf{e}_{1,k}$ and predicted error $\mathbf{e}_{2,k}$. Since the estimation error set $\mathbb{S}_{\mathbf{e}_{1,k}}$ can be generated by a Gaussian distribution with mean $\mathbf{0}$ and covariance matrix $\frac{\sigma^2_0}{SML}\mathbf{I}$ according to (\ref{es1}), we mainly focus on the generation of prediction error set $\mathbb{S}_{\mathbf{e}_{2,k}}$. Only for channel augmentation, we do not need to know the exact distribution of $\mathbf{e}_{2,k}$. It motivates us to use VAE which is a DL-based method for data generation. { The probabilistic framework of VAE provides a robust foundation for modeling complex data distributions, enabling the capture of intricate patterns and variations within the data \cite{VAE}. Moreover, the latent space representation offered by VAE facilitates the extraction of meaningful features, allowing for tasks such as data generation to be performed effectively. VAE’s generative modeling capabilities are particularly relevant in this context, as they align with the need for data generation and augmentation. Additionally, the regularization properties of VAE contribute to improved	generalization, while its inherent dimensionality reduction aids in feature extraction, further justifying its selection in the context of this research or application. These rationales collectively underscore the suitability of VAE technology for the specific objectives at hand, positioning it as a pertinent and valuable tool for the tasks and analysis to be undertaken.} As shown in Fig. \ref{VAE}, VAE utilizes the latent variables $\{\mathbf{z}^{(i)}\}_{i=1}^N$ to reconstruct the distribution of the observed prediction error vectors $\{\mathbf{e}^{(i)}\}_{i=1}^N$ accordingly (For simplicity, we assume that the complex prediction channel error has been changed into the real domain). Then, the probabilistic decoder is the likelihood function $p(\mathbf{e}^{(i)}|\mathbf{z}^{(i)})$ and naturally the posterior distribution $p(\mathbf{z}^{(i)}|\mathbf{e}^{(i)})$ is the encoder, which can be expressed as
	\begin{equation}\label{houyan}
		p(\mathbf{z}^{(i)}|\mathbf{e}^{(i)}) = \frac{p(\mathbf{e}^{(i)}|\mathbf{z}^{(i)})p(\mathbf{z}^{(i)})}{\int p(\mathbf{e}^{(i)}|\mathbf{z}^{(i)})p(\mathbf{z}^{(i)})d\mathbf{z}^{(i)}}.
	\end{equation}
	Unfortunately, the posterior distribution is intractable due to the complex computation of vector integrals in (\ref{houyan}). To address the problem, the variational inference is adopted which establishes a variational distribution $q(\mathbf{z}^{(i)}|\mathbf{e}^{(i)})$ to approximate the posterior $p(\mathbf{z}^{(i)}|\mathbf{e}^{(i)})$. To achieve the goal, $q(\mathbf{z}^{(i)}|\mathbf{e}^{(i)})$ is optimized to minimize the Kullback-Leibler (KL)
    divergence, i.e.,
    \begin{equation}\label{KL1}
    	\underset{q(\mathbf{z}^{(i)}|\mathbf{e}^{(i)})}{\mathop{\text{minimize}}}\,\quad \mathrm{KL}(q(\mathbf{z}^{(i)}|\mathbf{e}^{(i)})||p(\mathbf{z}^{(i)}|\mathbf{e}^{(i)})) 
    \end{equation}

    \emph{Theorem 1}: According to the properties of probability and KL divergence, the problem in (\ref{KL1}) can be transformed as
    \begin{equation}\label{KL2}
    	\underset{q(\mathbf{z}|\mathbf{e}^{(i)})}{\mathop{\text{minimize}}}\,\quad
    	-\mathbb{E}_{q(\mathbf{z}|\mathbf{e}^{(i)})}[\ln(p(\mathbf{e}^{(i)}|\mathbf{z}))]+ \mathrm{KL}(q(\mathbf{z}|\mathbf{e}^{(i)})||p(\mathbf{z})).
    \end{equation}

    \begin{IEEEproof} 
    Please refer to Appendix A.
    \end{IEEEproof}

\begin{figure}
    	\centering
    	\includegraphics [width=0.5\textwidth] {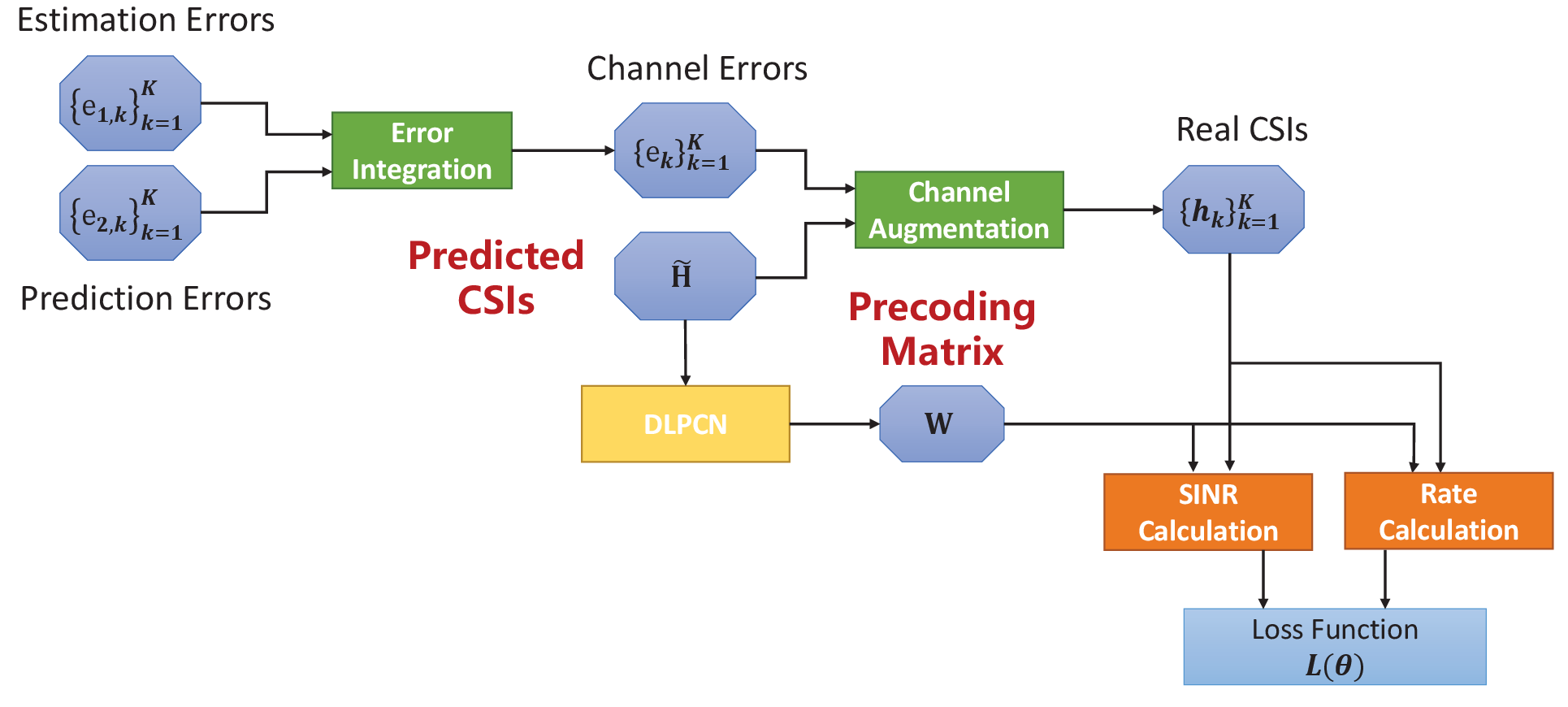}
    	\caption {The procedure of loss calculation.}
    	\label{loss}
    \end{figure}

    {After the transformation, the objective in (\ref{KL2}) is expressed in the form of known variables. It can be divided into two terms, the first term represents the reconstruction error $\Vert\mathbf{e}^{(i)}-\hat{\mathbf{e}}^{(i)}\Vert_2$, and the second term is a regularization, which ensures the generative capability of the model.} According to the central limit theorem, the standard Gaussian distribution is adopted for the prior distribution, i.e., $p(\mathbf{z})=\mathcal{N}(\mathbf{0},\mathbf{I})$, to facilitate sampling. Meanwhile, considering the integral calculation of KL divergence, Gaussian distribution is used for the variational distribution as $q(\mathbf{z}|\mathbf{e}^{(i)})=\mathcal{N}(\boldsymbol{\mu}_i^z, {(\boldsymbol{\sigma}_i^z)}^2)$, Then the optimization objective in (\ref{KL2}) can be reformulated as
    \begin{equation}\label{KL3}
    	\begin{aligned}
    		&-\mathbb{E}_{q(\mathbf{z}|\mathbf{e}^{(i)})}[\ln(p(\mathbf{e}^{(i)}|\mathbf{z}))]+ \mathrm{KL}(q(\mathbf{z}|\mathbf{e}^{(i)})||p(\mathbf{z})) \\
    	&	=\frac{1}{N}\sum_{i=1}^N-\frac{1}{2}\big(\ln {(\boldsymbol{\sigma}_i^z)}^2-{(\boldsymbol{\sigma}_i^z)}^2-(\boldsymbol{\mu}_i^z)^2\big)\\
    	&+\frac{1}{N}\sum_{i=1}^N\Vert\mathbf{e}^{(i)}-\hat{\mathbf{e}}^{(i)}\Vert_2^2,\\
    	\end{aligned}
    \end{equation}
    which is defined as the loss function of VAE. Based on (\ref{KL3}), two MLPs defined as mean encoder and variance encoder as illustrated in Fig. \ref{VAE} are constructed to generate the mean $\boldsymbol{\mu}_i^z$ and variance ${(\boldsymbol{\sigma}_i^z)}^2$, respectively. Each MLP has two FC layers with $3N_\text{in}$ and $N_\text{in}$ neurons, while $N_\text{in}$ is the number of input prediction error samples. Next, a latent variable $\mathbf{z}^{(i)}$ is sampled from $q(\mathbf{z}^{(i)}|\mathbf{e}^{(i)})=\mathcal{N}(\boldsymbol{\mu}_i^z,{(\boldsymbol{\sigma}_i^z)}^2)$. Finally, a decoder constructed of MLP is designed to reconstruct the sample $\hat{\mathbf{e}}^{(i)}$. After training, we keep the decoder to generate a large number of samples by sampling the latent variables from $p(\mathbf{z}^{(i)})=\mathcal{N}(\mathbf{0},\mathbf{I})$. By this way, we can enrich the number of samples in $\mathbb{S}_{\mathbf{e}_{2,k}}$. Subsequently, we can obtain the channel error set $\mathbb{S}_{\mathbf{e}_{k}}$ as
    \begin{equation}\label{setg}
    	\mathbb{S}_{\mathbf{e}_{k}}= \{\mathbf{e}_{1,k}+\boldsymbol{\xi}_k\cdot\mathbf{e}_{2,k}, \forall  \mathbf{e}_{1,k}\in\mathbb{S}_{\mathbf{e}_{1,k}},\forall\mathbf{e}_{2,k}\in\mathbb{S}_{\mathbf{e}_{2,k}} \},
    \end{equation}
    which is used as an auxiliary set for loss training. In this way, the outage constrained problem in (24) can be addressed by DLPCN. To make it more clear,  Fig. \ref{loss} concludes the whole procedure of loss calculation.

\begin{figure}
    	\centering
    	\includegraphics [width=0.5\textwidth] {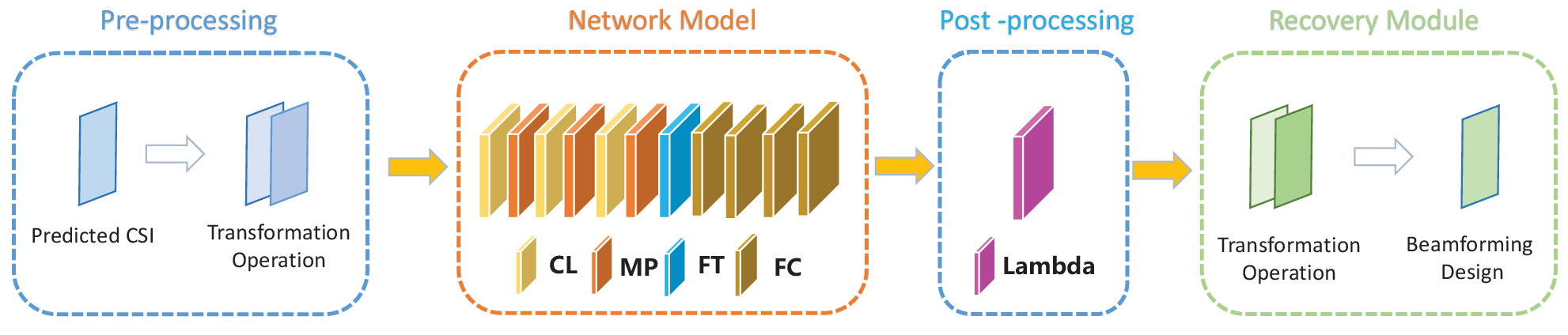}
    	\caption {The network structure for DLPCN in LEO satellite IoT.}
    	\label{network_2}
    \end{figure}

    Eventually, we introduce the DLPCN as illustrated in Fig. \ref{network_2}. Since the structure of recovery module is similar to that in DLPDN, we mainly discuss the pre-processing, network model and post-processing.

    \emph{1) Pre-processing:} To satisfy the input requirement of DNN, we use $\zeta(\cdot)$ to transform $\widetilde{\mathbf{H}}$ into the real domain, that is,
    \begin{equation}\label{realpart2}
    	\zeta(\widetilde{\mathbf{H}})=\bigg[\begin{array}{ll}
    		\Re\{\widetilde{\mathbf{H}}\} \\
    		\Im\{\widetilde{\mathbf{H}}\} \\
    	\end{array}
    	\bigg]\in\mathbb{C}^{2M\times K}.
    \end{equation}
    Then we use $\zeta(\widetilde{\mathbf{H}})$ as the input of DLPCN.

    \emph{2) Network Model:} As shown in Fig. \ref{network_2}, three CL layers with ReLU activation are adopted to exact the features of predicted CSI. Then, a flatten (FT) layer is adopted after the last CL layer to reshape the data in the formation of the suitable input of FC layer. Finally, four FC layers with $8MK$, $4MK$, $2MK$, and $2MK$ neurons are constructed to calculate the result, which is represented by $\mathbf{w}_{\text{real}}$. We use (\ref{OP3obj}) for the network training.

    \emph{3) Post-processing:}  {  According to the constraint (34b), since the sum power of the precoding vectors, i.e., the power of DNN outputs $\sum_{k=1}^{K}\vert\vert\mathbf{w}_k\vert\vert^2=\vert\vert\mathbf{w}_\text{total}\vert\vert^2$ is upper bounded by the total power $P_2$, we employ a DNN structure to address the total power constraint by constructing a Lambda-layer after the output layer.} Specifically, the Lambda layer includes the  following two steps:

\begin{enumerate}
	\item Normalize the output vector of DNN, i.e., $\mathbf{w}_{\text{real}}^\text{norm} = \frac{\mathbf{w}_{\text{real}}}{\vert\vert\mathbf{w}_{\text{real}}\vert\vert}$;
	\item Multiply the normalized output vector $\mathbf{w}_{\text{real}}^\text{norm}$ by the total transmit power $P_2$, i.e., $\mathbf{w}_{\text{real}}=\sqrt{P_2}\cdot\mathbf{w}_{\text{real}}^\text{norm}$
\end{enumerate}

    Therefore, by using the proposed DLPCN, it is likely to generate high-performance multibeam precoding vectors based on predicted CSI obtained with the proposed DLPDN. { Finally, the proposed joint channel prediction and multibeam precoding scheme is summarized in \textbf{Algorithm 1}.}

    \begin{breakablealgorithm}
    	\caption{: Joint Channel Prediction and Multibeam Precoding for LEO Satellite IoT}
    	\label{alg1}
    	
    	\begin{algorithmic}[1]
    		\STATE{\textbf{Offline Training 1: }}
    		
    		\STATE{\textbf{~~~~Input:}  Estimated CSI dataset ${\{\hat{\mathbf{H}}_k}\}_{k=1}^K$}
    		
    		\STATE{\textbf{~~~~Output:}  Trained DLPDN model}
    		
    		\STATE{~~~~ Initialize the DLPDN parameters.}
    		\STATE{~~~~ \textbf{For} 1 : epoches \textbf{do}}
    		\STATE{~~~~~~~~Obtain the output of the network $\{\Tilde{\mathbf{H}}_k\}_{k=1}^K$.}
    		\STATE{~~~~~~~~Calculate the loss function, i.e., the NMSE metric. }
    		\STATE{~~~~~~~~Update the network parameters by Adam optimizers.}
    		\STATE{~~~~ \textbf{End For}}
    		\STATE{~~~~Output the best DLPDN model via a validation set.}
    		\STATE{\textbf{Offline Training 2: }}
    		
    		\STATE{\textbf{~~~~Input:}  Predicted CSI dataset ${\{\tilde{\mathbf{H}}_k}\}_{k=1}^K$}
    		
    		\STATE{\textbf{~~~~Output:}  Trained DLPCN model}
    		
    		\STATE{~~~~ Initialize the DLPCN parameters.}
    		\STATE{~~~~ \textbf{For} 1 : epoches \textbf{do}}
    		\STATE{~~~~~~~~Obtain the output of the network $\{\mathbf{w}_k\}_{k=1}^K$.}
    		\STATE{~~~~~~~~Calculate the loss function in (35).}
    		\STATE{~~~~~~~~Update the network parameters by Adam optimizers.}
    		\STATE{~~~~ \textbf{End For}}
    		\STATE{~~~~Output the best DLPCN model via a validation set.}
    		
    		\STATE{\textbf{Online Joint Channel Prediction and Multibeam Precoding:}}
    		
    		\STATE{\textbf{~~~~Input:}  Past estimated CSI $\hat{\mathbf{H}}$, trained DLPDN model, trained DLPCN model}
    		
    		\STATE{\textbf{~~~~Output:}  The predicted CSI matrix $\tilde{\mathbf{H}}$, the precoding matrix $\mathbf{W}$.}
    		
    		\STATE{~~~~Initialize the network parameters according to the trained DLPDN model.}
    		\STATE{~~~~Obtain the output of the network as the predicted CSI matrix $\tilde{\mathbf{H}}$.}
    		\STATE{~~~~Initialize the network parameters according to the trained DLPCN model.}
    		\STATE{~~~~Obtain the output of the network as the precoding matrix $\mathbf{W}$.}

    	\end{algorithmic}
    \end{breakablealgorithm}

 \subsection{Complexity Analysis}
    Herein, we analyze the computational complexity of the two proposed DL-based networks, i.e., DLPDN and DLPCN, respectively. {Firstly, according to \cite{complex}, the computational complexity for $l$-th CL layer is given by $O( C_l^K C_l^{in} C_l^{out} F_l)$, where $C_l^K$, $C_l^{in}$, $C_l^{out}$, $F_l$ are the area of kernel for the $l\text{-th}$ CL layer, the number of input channel, the number of output channel, and the output feature map size, respectively. Secondly, the computational complexity of one LSTM layer can be expressed as $O(4F_{l-1}F_{l}T_{l}^{ne})$, where $T_l^{ne}$ denotes the number of neurons for $l$-th LSTM layer. Thirdly, the complexity of $l$-th FC-layer is $O(F_{l-1}F_{l})$. While the network model of DLPDN is composed of three CL layers, two LSTM layers, and one FC layer, the total computational complexity of DLPDN is $O(\sum_{l=1}^{3} C_l^K C_l^{in} C_l^{out} F_l+\sum_{4}^{5} 4F_{l-1}F_{l}T_{l}^{ne}+F_{5}F_{6})$. Similar to DLPDN, DLPCN has three CL layers and four FC layers, so the computational complexity of DLPCN is $O(\sum_{l=1}^{3} C_l^K C_l^{in} C_l^{out} F_l+\sum_{4}^{7}F_{l-1}F_{l})$.}

\section{Simulation Results}
    In this section, we present extensive simulation results to validate the effectiveness of the proposed DL-based joint channel prediction and multibeam precoding scheme for LEO satellite IoT. The main simulation system parameters for LEO satellite IoT is set up according to 3GPP TR 38.821, which is summarized in Table I.
\begin{table}
	\small
	\centering
	\caption{Main Simulation System Parameters For LEO Satellite IoT precoding scheme}\label{Simulation}
	\begin{tabular}{|c|c|}
		\hline
		Parameter & Value\\\hline\hline
		Satellite orbit & LEO \\\hline
		Frequency band & L/S/C  \\\hline
		Carrier frequency $f_c$& 5 GHz  \\\hline
		Altitude of orbit $d_0$ & 1000 km \\\hline
		{LEO satellite velocity $v_0$} &{ 7.35 km/s}\\\hline
		{Doppler shift of LEO satellite $\nu_{k}^{\text{Sat}}$} &{120 kHz}\\\hline
		{Maximum doppler shift of IoT devices $\nu_{k,l}^{\text{Dev}}$} &{20 Hz}\\\hline
		{Minimum time delay $\tau_{k}$} &{10 ms}\\\hline
		Carrier bandwidth $B$  & 25MHz  \\\hline
		Satellite antenna gain $\omega_k$  & 17 dBi  \\\hline
		{IoT device receive antenna gain $G_k$}  & {3 dBi}  \\\hline
		Noise temperature $T$ & 300 K \\\hline
		Boltzmann's constant $\kappa$ & 1.38 $\times 10^{-23} $ J/m \\ \hline
		Rain fading mean $\mu_r$& -2.6 dB \\\hline
		Rain fading variance $\sigma_r^2$ &  1.63 dB \\\hline
		3dB angle & $0.4^{\circ}$ \\\hline
		Rician factor $\lambda$ & 5 \\\hline
		{Noise variance  $\sigma_0^2$} & {$-106$ dBm \cite{noise}}\\ \hline
		{Pilot sequence power $P_1$}  &{$10$ dBw} \\ \hline
		Number of antennas $M$ & $16$\\ \hline
		{Number of IoT devices $K$} & {$16$}\\\hline
		Weight of IoT devices $\alpha_k$ & $1$  \\ \hline
		Outage probability $p_k$ & $0.05$ \\ \hline
		Minimum required SINR threshold $\gamma_k$ & $0$ dB \\ \hline
		Total transmit power $P_2$  &$10$ dBw \\ \hline
	\end{tabular}
\end{table}	

\subsection{Channel Prediction}
    At the channel prediction part, the structure of DLPDN is constructed as illustrated in Fig. \ref{network1} for simulation. The estimated CSI dataset is generated according to (\ref{es1}){\footnote{Actually, in practical scenarios, the accuracy of estimated CSI is affected by many factors, including the length and power of pilot sequences, which have been taken into account in Eq. (8). Therefore, the generated dataset matches the practical scenarios to a certain extent.}}, the number of training samples and testing samples are set as 16000 and 4000, respectively, and the input step size $w_{\text{step}}$ is set as 4. {For the parameter settings of the network model, the number of the filters of each CL layer are chosen as 8, 4, 2, and the filter size is set as (5,1) (3,1) (3,1), respectively.} To train the DLPDN, we adopt the mini-batch SGD algorithm based on the Adam optimizer. The initial learning rate is set as 0.001 and the batch size is set as 128. Moreover, the number of training epochs is set as 500. We use normalized mean square error (NMSE) to measure the accuracy of channel prediction, which is defined as $\frac{\Vert\hat{\mathbf{H}}-\widetilde{\mathbf{H}}\Vert}{\Vert\hat{\mathbf{H}}\Vert}$.
    \begin{figure}
	    \centering
     	\includegraphics [width=0.5\textwidth] {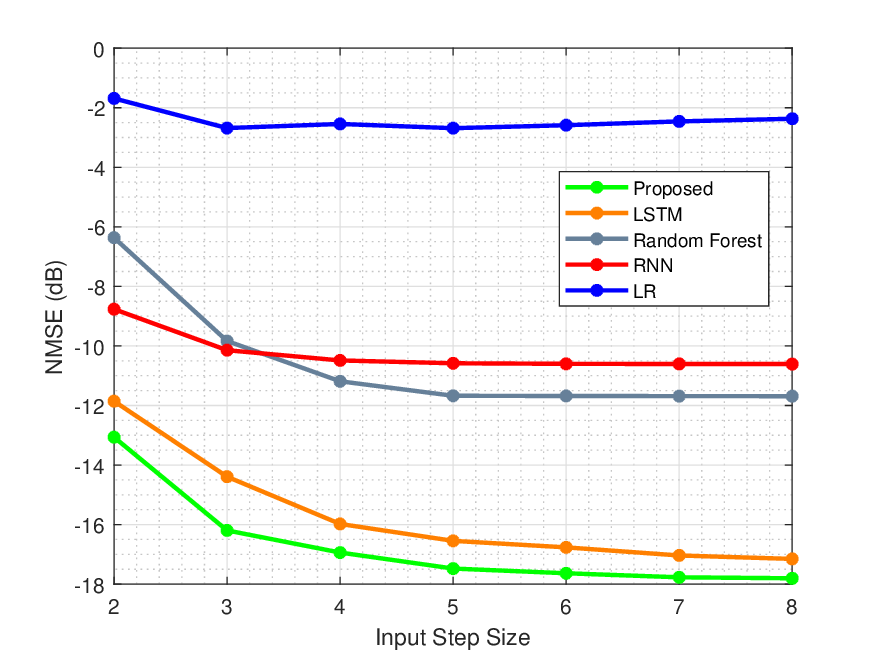}
    	\caption {The NMSE performance comparisons of channel prediction schemes under various input step sizes.}
    	\label{NMSE_IS}
    \end{figure}

    {First, we compare the performance of the proposed DL-based channel prediction scheme with four baseline schemes under various input step sizes as illustrated in Fig. \ref{NMSE_IS}. One is linear regression (LR) method proposed in \cite{conv} is a basic method for channel prediction.} An LSTM scheme is proposed in \cite{ml2} whose network is totally composed of the LSTM layers, a RNN scheme is proposed in \cite{RNN} which replaces the LSTM layers with the classical RNN layers, and the last is the random forest algorithm \cite{RF}. { It can be seen that our proposed scheme has the lowest NMSE among four schemes owing to the CL layers which can effectively extract the spatial features of the estimated CSI, while the RNN and LSTM scheme only extract the temporal features of CSI. Correspondingly, the added convolution layer may increase the computational complexity of the DL scheme.} Moreover, the NMSE decreases with the increment of input step size, so it makes sense to choose an appropriate input step size to balance channel prediction accuracy and computational resource consumption. {Furthermore, it can be seen in Fig. 9 that LR achieves the worst performance due to complex LEO satellite channels. However, the DL technique can adapt to the variation of LEO satellite channels and accurately predict channel state information. Therefore, the DL technique is effective for channel prediction in LEO satellite IoT.}
    \begin{figure}
     	\centering
    	\includegraphics [width=0.5\textwidth] {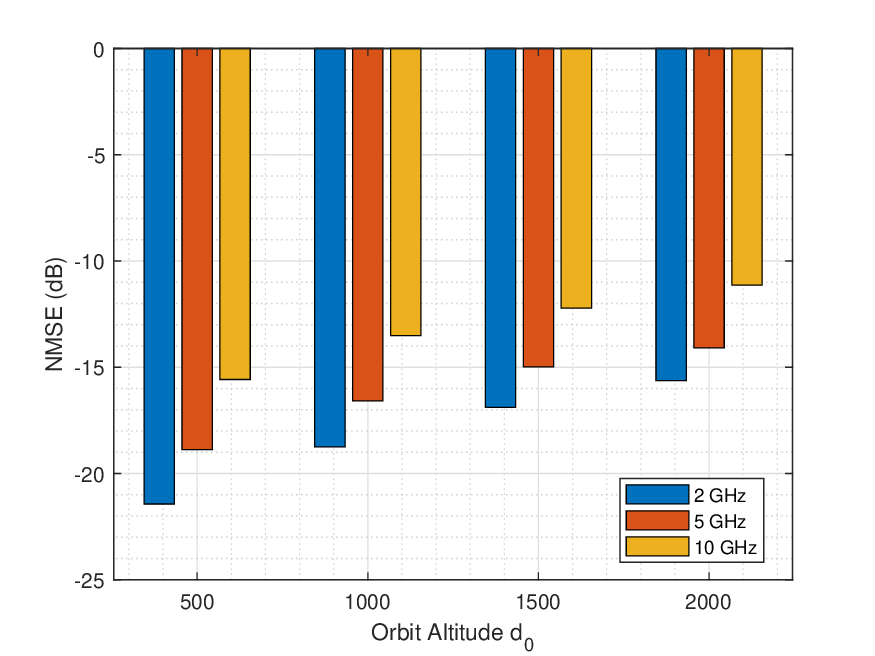}
    	\caption {The NMSE performance of the proposed scheme with different carrier frequencies $f_c$ under various orbit altitudes $d_0$.}
     	\label{NMSECA}
    \end{figure}

    Then, we explore the impact of carrier frequency $f_c$ and orbit altitude $d_0$ on the performance of channel prediction. It can be seen from Fig. \ref{NMSECA} that the prediction accuracy decreases as the carrier frequency and orbit altitude increase. This is because both the increment of the orbit altitude and carrier frequency may weaken the auto-correlation of the channel by increasing the free-space path loss in (\ref{g_k}). Therefore, a satellite operating at a lower orbit altitude and a lower carrier frequency can improve the quality of channel prediction effectively. For instance, very low earth orbit is now applied for satellite communications.

\subsection{Multibeam Precoding}
    Now let we focus on the performance evaluation of multibeam precoding. The network structure is constructed as Fig. \ref{network_2}. {For the parameters settings of DLPCN, the numbers of the filters are 8, 4, and 2, respectively. The filter size and stride are chosen as (5,3), (5,1), (3,1) and one.} Moreover, for efficient NN training, batch-normalization technique is applied after each FC layer except the last layer to accelerate the training rate. Besides, the NN parameters are initialized according to the Xavier initialization \cite{Xavier}. We notice the proposed DLPCN scheme is based on a kind of unsupervised learning approach as we explained before, which only involves predicted CSIs and channel errors during the training procedure. To this end, we generate predicted CSI set with size  $10^6$.
     For channel error set, we generate $10^4$ samples for channel prediction error set, and $10^4$ samples for channel estimation error set. Then, a channel error set of size $10^8$ is constructed according to (\ref{setg}). Next, the validation split factor is set 0.2 for the proposed training method, and the NN parameters are optimized by the mini-batch SGD algorithm based on the Adam optimizer \cite{Adam}. To be more specific, the initial learning rate and the batch size are set as 0.001 and 1024, respectively. In each training batch, every training sample is augmented with $10^4$ channel errors which are randomly selected from the channel error set. Finally, the maximum number of epochs is 1000. To improve the training efficiency, the early stopping technique is adopted with iterations 20 for the validation loss with an accuracy of 0.0001 to prevent overfitting resulting from overtraining, and ReduceLROnPlateau technique is used to update the learning rate of Adam for a better loss performance with patient 10 and decreasing factor of 0.3.

    For testing, we use $10^3$ testing samples which are generated as training samples and each sample is augmented with $10^4$ channel errors to evaluate the SINR performance, and the outputs of NN are used to reconstruct the precoding vectors and calculate the WSR performance. The whole model is trained with the open-source machine learning framework Tensorflow 2.11.0 on GPU NVIDIA GeForce RTX 3060.

 \begin{figure}
 	\centering
 	\subfigure[Training loss]{\includegraphics[width=1\linewidth]{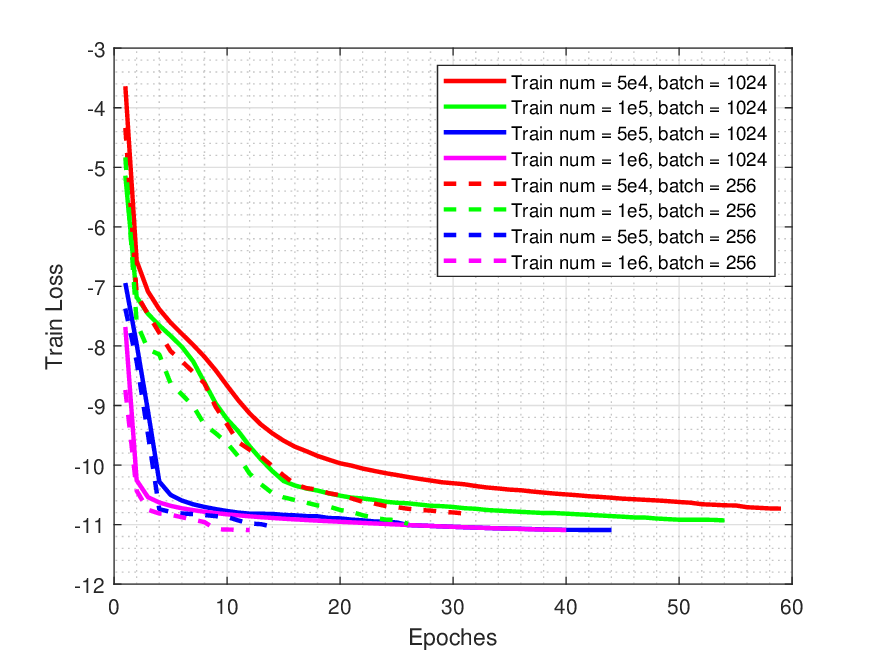}}
 	\subfigure[Validation loss]{\includegraphics[width=1\linewidth]{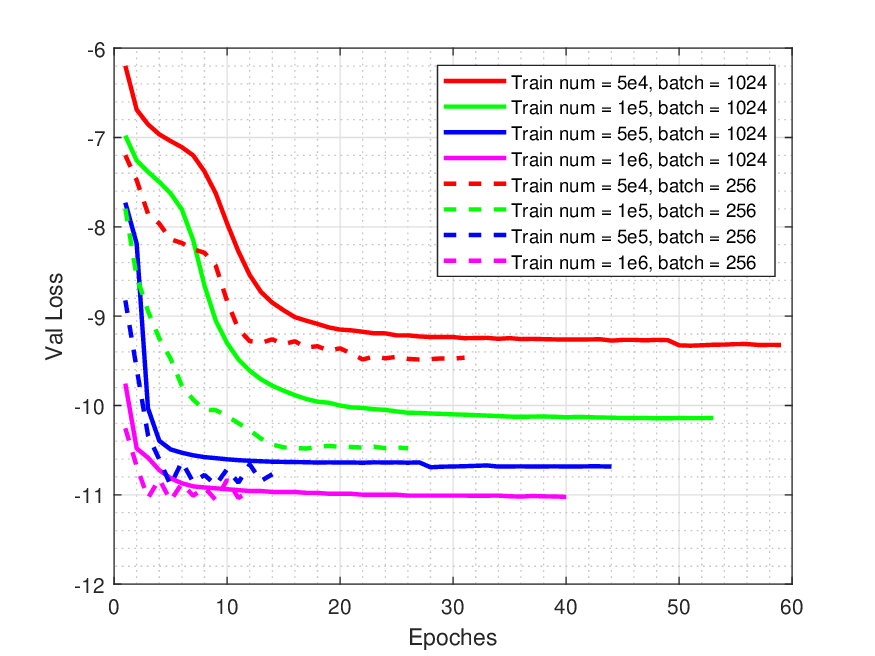}}
 	\caption {Training loss and validation loss for the proposed algorithm under different numbers of training examples and batchsize.}
 	\label{its}
 \end{figure}

    Firstly, we investigate the training performance of the proposed DL-based multibeam precoding scheme with different numbers of training samples. It can be seen from Fig. \ref{its}(a) and Fig. \ref{its}(b) that both the training loss and validation loss decrease gradually until convergence with the increment of epochs, which illustrates that our dataset is well distributed and the designed NN structure is effective. Furthermore, comparing Fig. \ref{its}(a) and Fig. \ref{its}(b), it is found that the validation loss at convergence is much larger than the training loss when there are a small number of training samples. For instance, when the number of training samples is $5\times10^4$, the validation loss is $-9.3$, much higher than the training loss $-10.8$. It means that the proposed NN model is overfitting, which can be solved by increasing the number of training samples. Moreover, it is also shown that with the increase of training examples, the gap between the validation loss and training loss becomes smaller, which effectively mitigates the impact of overfitting. { Besides, the simulation results shows that more epoches are needed to converge to the minimum validation loss with a large number of batchsize. However, when the batchsize is small, the validation loss begins to oscillate as the number of training samples increases, resulting in a slow convergence speed or even misconvergence.}

     \begin{figure}
    	\centering
    	\includegraphics [width=0.5\textwidth] {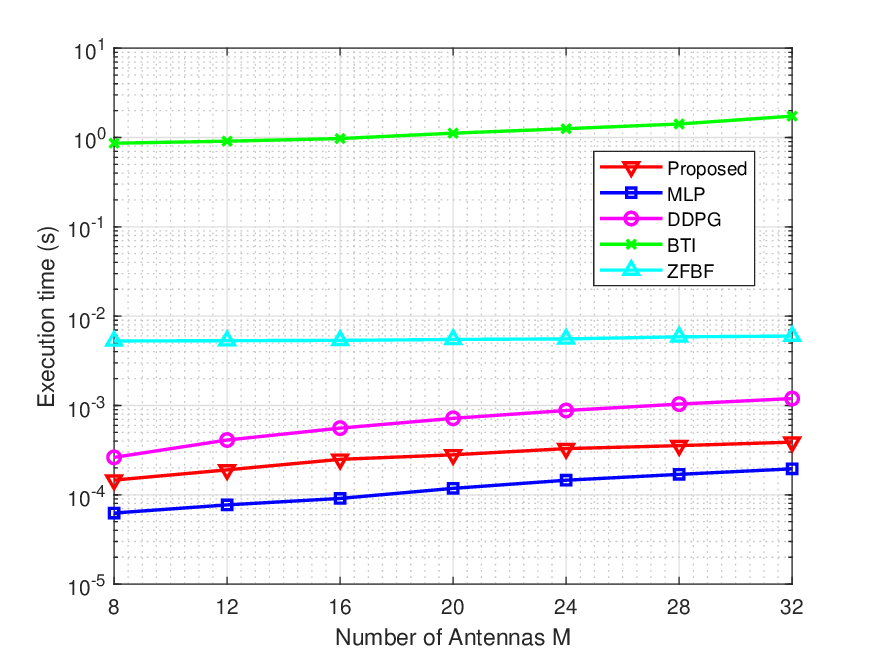}
     	\caption{Execution time comparisons of different schemes under various numbers of antennas $M$.}
        \label{TM}
    \end{figure}

   \begin{figure}
    	\centering
    	\includegraphics [width=0.5\textwidth] {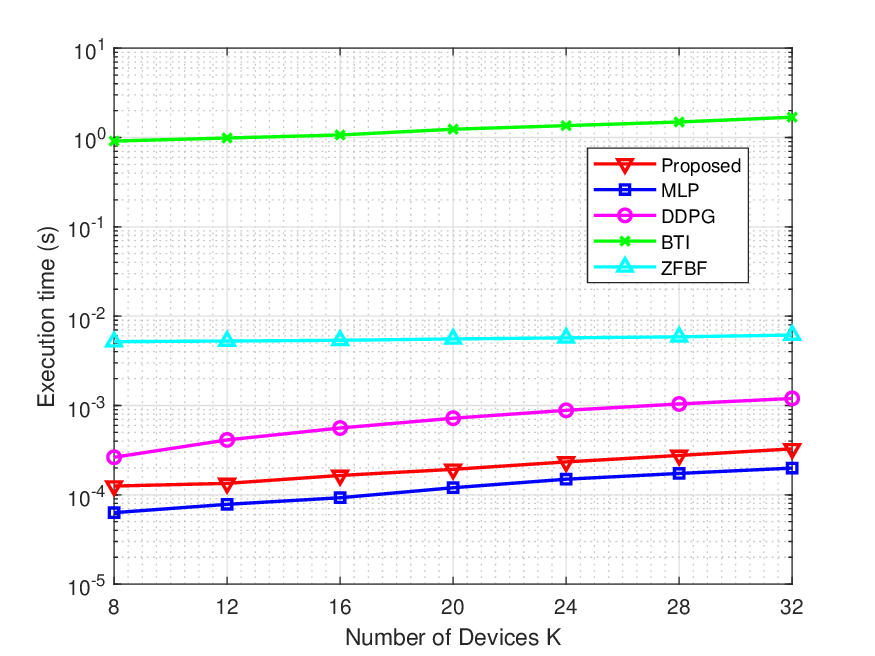}
    	\caption{Execution time comparisons of different schemes under various numbers of devices $K$.}
    	\label{TK}
    \end{figure}

    Next, we study the execution time of the proposed DL-based multibeam precoding scheme under different numbers of antennas $M$ and different numbers of devices $K$. It is important for the proposed scheme to have a short execution time to adapt the rapidly changing channels in LEO satellite IoT. To verify the time efficiency, {we use four baseline schemes for comparison, including the MLP scheme proposed in \cite{ml8}, the Deep Deterministic Policy Gradient (DDPG) scheme in \cite{ddpg}, the Bernstein-type inequality (BTI) scheme in \cite{BTI}, and the zero-forcing beamforming (ZFBF) scheme in \cite{ZFBF}.} To be more specific, MLP is also a DL-based scheme whose network only consists of FC layers, {DDPG is a classical algorithm of deep reinforce learning (DRL),} BTI is a numerical scheme with high performance and ZFBF is also a numerical scheme with low complexity. These baseline schemes will continue to be adopted in the subsequent simulations. Besides, to prevent serendipity, we adopt the average execution time as the metric to measure the computational complexity performance. In Fig. \ref{TM} and \ref{TK}, we show the execution time of the four schemes under different numbers of antennas $M$ and numbers of devices $K$, respectively. The proposed scheme needs slightly longer execution time than the MLP scheme, but much shorter time than the ZFBF scheme and BTI scheme. This is because our proposed scheme has more NN layers than MLP scheme, and numerical schemes need to perform multiplication and inversion of high-dimensional matrix in each iteration, which require long time to complete the iterative process, especially for BTI scheme. Therefore, our proposed scheme is time-competitive among the schemes.
    \begin{figure}
    	\centering
    	\includegraphics [width=0.5\textwidth] {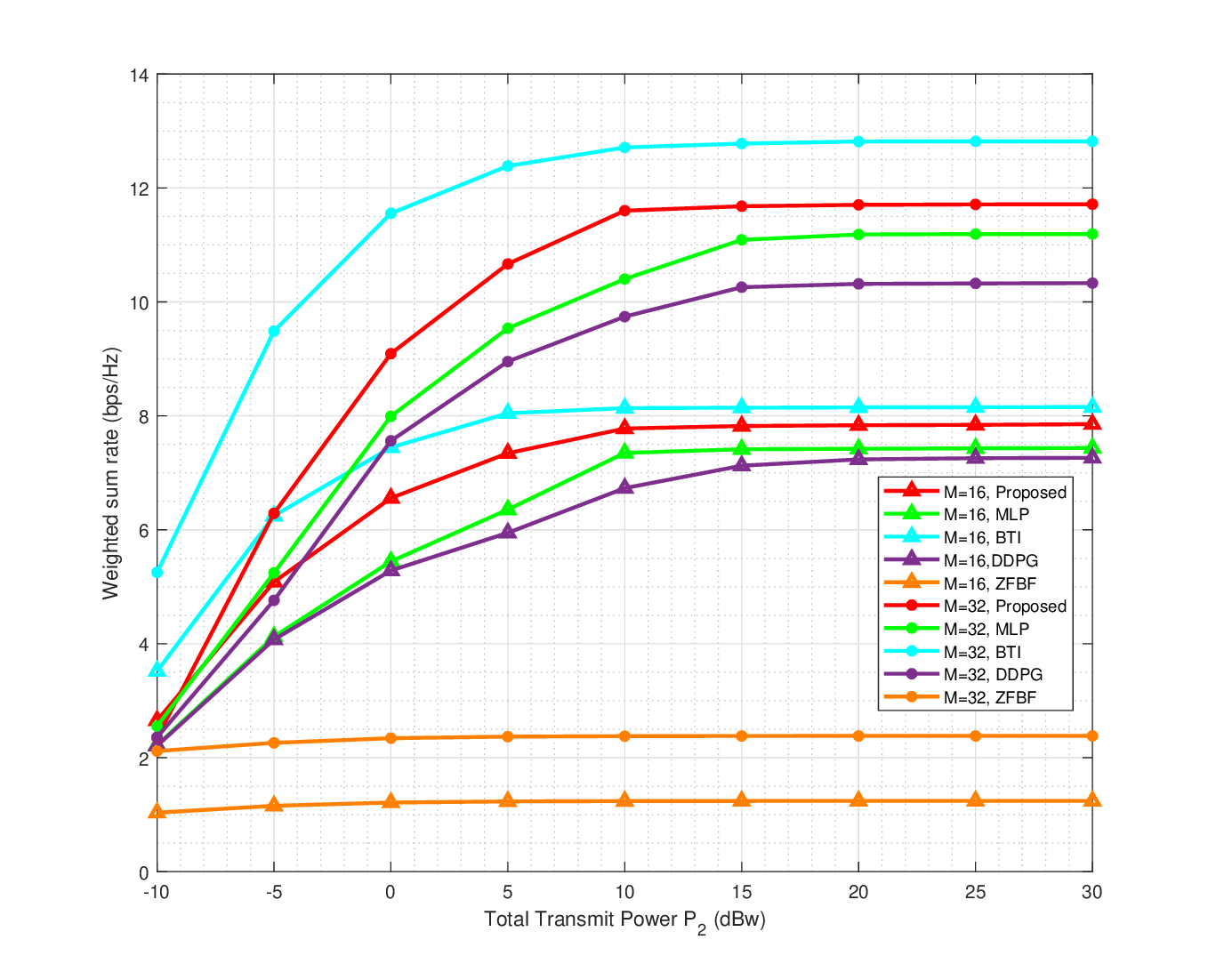}
    	\caption {The WSR performance comparisons of different schemes under various total transmit powers $P_2$.}
    	\label{M_P}
    \end{figure}

    Due to limited power supply, power consumption is also an important metric for LEO satellite IoT. In Fig. \ref{M_P}, we investigate the impacts of the total transmit power $P_2$ under different numbers of antennas $M$ on the WSR performance of the four schemes.  It is seen that with the increment of the total transmit power, the WSR first increases, and then saturates. Furthermore, the WSR improves as the number of antennas $M$ increases, since higher antenna array gains are exploited. Therefore, it is possible to increase the WSR performance of LEO satellite IoT by adding the number of antennas. It is also found that the WSR performance of the proposed scheme is slightly worse than BTI scheme but is much better than other two schemes. However, the execution time of BTI scheme is much larger than our proposed scheme as mentioned before. To this end, our proposed scheme is appealing to LEO satellite IoT.
    \begin{figure}
    	\centering
    	\includegraphics [width=0.5\textwidth] {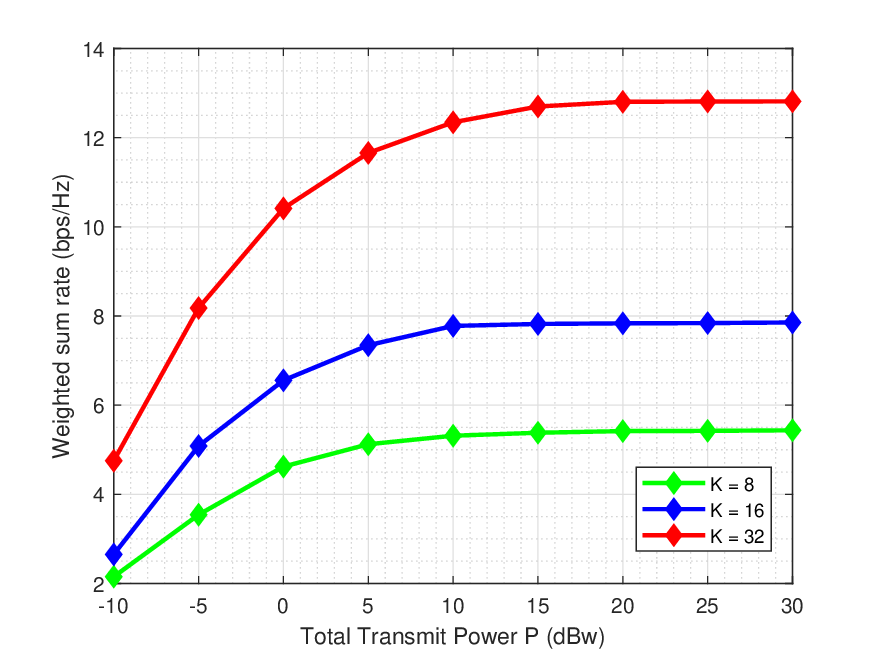}
    	\caption {The WSR performance of the proposed scheme with different number of devices $K$ under various total transmit powers $P_2$.}
    	\label{KP}
    \end{figure}

    Further, we investigate the WSR performance of the proposed multibeam precoding scheme with different numbers of devices under various total transmit powers. As illustrated in Fig. \ref{KP}, the WSR first increases and then comes to convergence as total transmit power $P_2$ increases. Yet, for a given total transmit power, the saturated WSR increases with the increment of the number of devices $K$. Hence, the proposed scheme can admit more devices, which is of great importance for LEO satellite IoT.
    \begin{figure}
	    \centering
    	\includegraphics [width=0.5\textwidth] {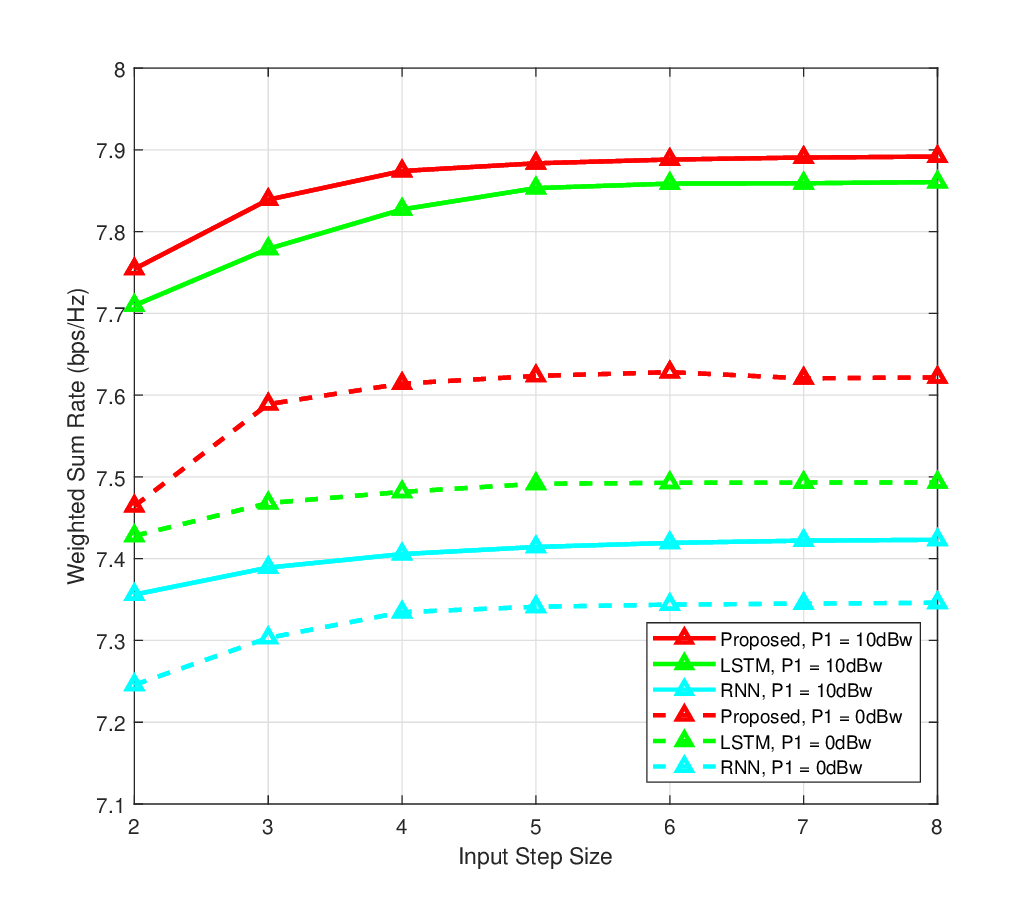}
    	\caption {The WSR performance comparisons of channel prediction schemes and the power of pilot sequences in channel estimation under various input step sizes.}
    	\label{RIS}
    \end{figure}

    It can be found in Eq. (8) that the transmit power for channel estimation $P_1$ affects the estimation error, while the prediction techniques influence the predicted error. Therefore, we investigate the impacts of different channel prediction schemes and the power of pilot sequences under various input step sizes on the WSR performance. It can be seen that the estimation error and prediction error have joint impacts on the WSR performance. Therefore, we can improve the system performance from two perspectives. On the one hand, the estimation error is decreased by increasing the power of pilot sequences. On the other hand, the prediction error is decreasing by adopting advanced prediction methods. Besides, it can be seen from Fig. \ref{RIS} that the proposed scheme has the best WSR performance among three schemes. This is because the proposed scheme has the strongest channel prediction capability compared to the other two schemes. Moreover, increasing the input step size can improve the WSR performance due to more accurate predicted CSI. However, the increment of the input step size may lead to a higher computational complexity which will result in more time spending on the network training. Therefore, we should choose an appropriate input step size for LEO satellite IoT.

     \begin{figure}
    	\centering
    	\includegraphics [width=0.5\textwidth] {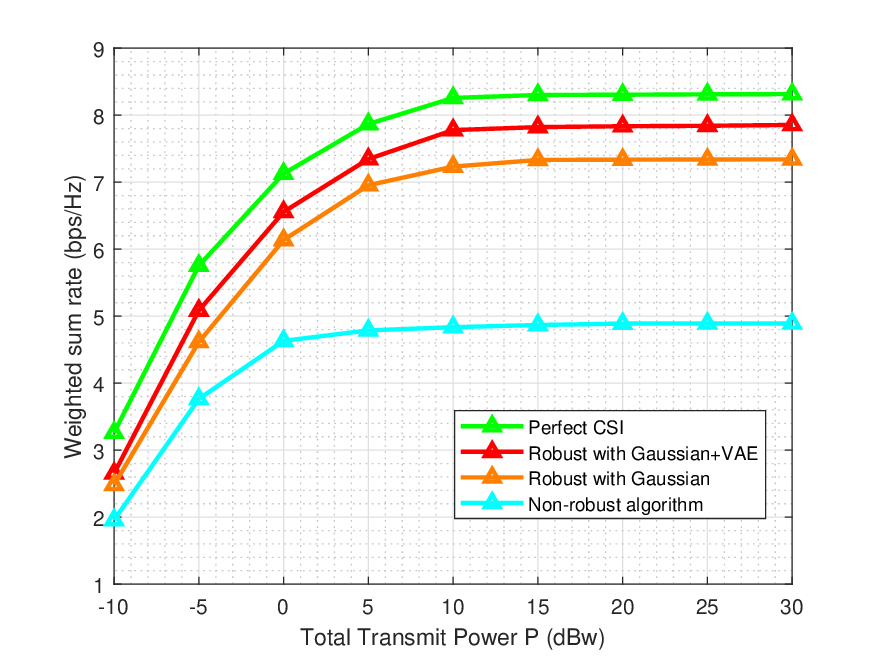}
    	\caption {The WSR performance comparisons of robust schemes with different channel error generation methods under various total transmit powers $P_2$.}
    	\label{Robust}
    \end{figure}

    To verify the effectiveness of introduced VAE in channel error set, we compare the proposed robust scheme with a non-robust scheme and a robust scheme whose error set is generated based on Gaussian distribution \cite{ml8}. It is shown in Fig. \ref{Robust} that robust schemes can significantly improve the WSR performance compared to the non-robust scheme. Moreover, the WSR is also increased by applying VAE in channel error set generation compared to the scheme whose error set is completely generated from Gaussian distribution. Therefore, our proposed channel augmentation method with VAE is competitive in robust schemes.    {Besides, it can be seen from the simulation result that the performance of the proposed algorithm with estimated CSI is close to that with perfect CSI. Hence, the proposed algorithm is highly robust to channel estimation error, which is appealling to LEO satellite IoT with high-dynamic environments.}

    \begin{figure}
    	\centering
    	\includegraphics[width=0.5\textwidth]{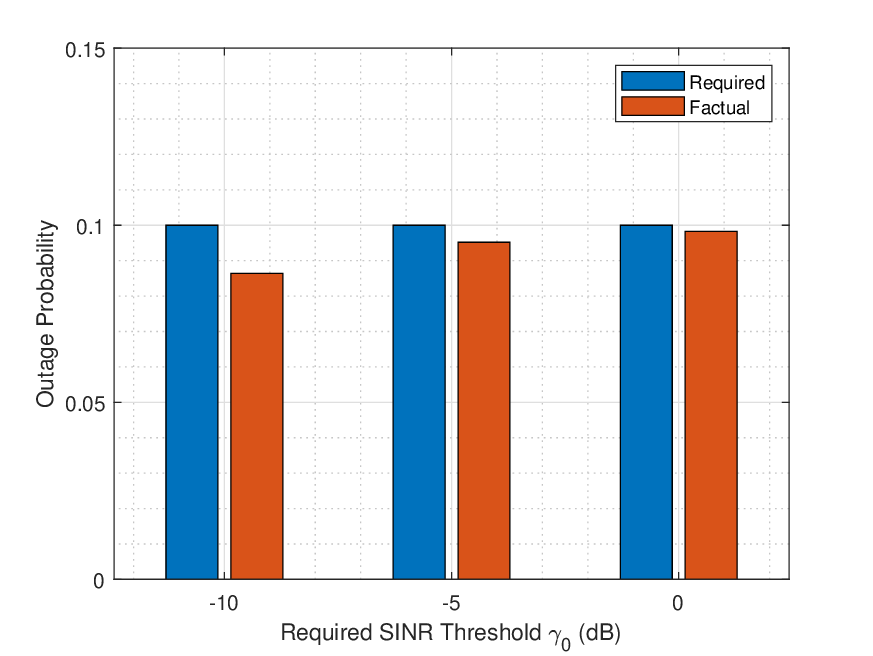}
    	\caption{The performance comparisons of outage probability under various required SINR threshold $\gamma_0$.}
    	\label{prb}
    \end{figure}

      \begin{figure}
     	\centering
     	\includegraphics[width=0.5\textwidth]{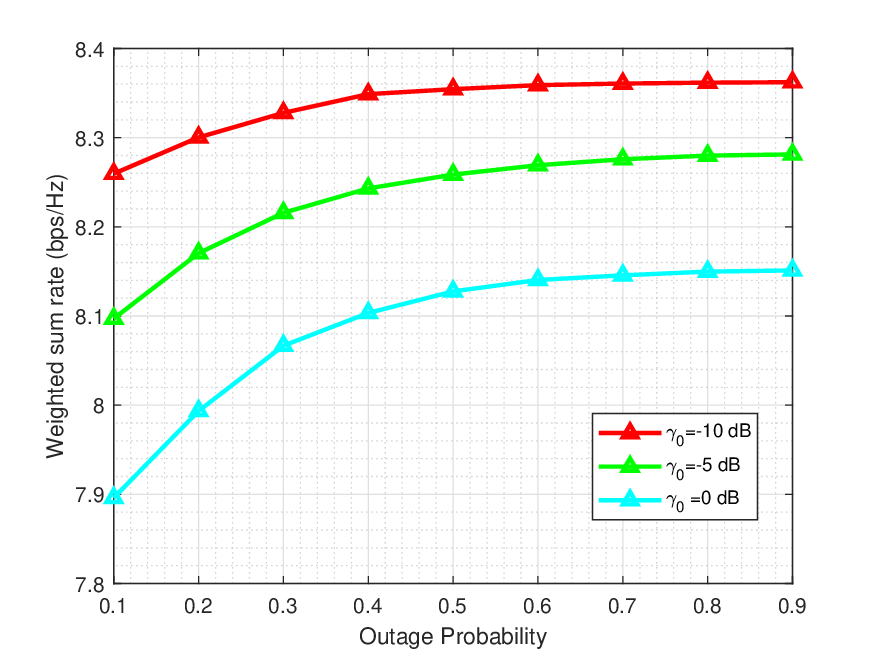}
        \caption{The WSR performance of the proposed scheme under various outage probabilities with different SINR thresholds $\gamma_0$.}
        \label{PSINR}
     \end{figure}

    Noticing that the outage probability constraint is involved in the calculation of loss function, it is essential to verify whether the factual outage probability satisfies the outage probability constraint. As shown in Fig. \ref{prb}, the factual outage probability is smaller than the required one, which illustrates that the proposed scheme produces a proper outage probability. Furthermore, it can be seen from Fig. \ref{prb} that the factual outage probability increases with the increment of SINR requirement. This is because higher SINR requirement needs more transmit power, resulting in the increment of outage probability for IoT devices.

    Finally, we examine the effect of SINR outage probability on the proposed DL-based multibeam precoding scheme. As can be observed in Fig. \ref{PSINR}, the WSR grows with the increment of the SINR outage probability. This corresponds with the fact that a lower outage probability requires more power to satisfy the SINR threshold. Besides, it can be found in Fig. \ref{PSINR} that a higher SINR threshold also leads to a lower WSR, because a higher SINR threshold requires more power to meet the SINR requirement. Therefore, it is important to choose a proper SINR outage probability in the multibeam precoding design to achieve a balance between design conservatism and service fidelity.

\section{Conclusion}
In this paper, we presented a joint channel prediction and multibeam precoding scheme for LEO satellite IoT under fast time-varying propagation environments. In particular, we first proposed a DL-based channel prediction scheme DLPDN for downlink CSI acquisition in the context of fast mobility of LEO satellite. Then, to realize the critical IoT device applications in LEO satellite IoT that require higher performance of communications, the predicted CSI was utilized to design a DL-based robust multibeam precoding scheme DLPCN by maximizing the WSR subject to outage constraints on the SINR of IoT devices. Finally, extensive simulation results validated the robustness and effectiveness of the proposed scheme, which successfully demonstrated that our proposed scheme is appealing to LEO satellite IoT.

\begin{appendices}
\section{The Proof of Theorem 1}
    The optimization objective in (\ref{KL2}) can be obtained according to (\ref{KL1}) by the following steps:
  \begin{equation}
  	\begin{aligned}
  		&\mathrm{KL}(q(\mathbf{z}^{(i)}|\mathbf{e}^{(i)})||p(\mathbf{z}^{(i)}|\mathbf{e}^{(i)}))\\
  		=&\int q(\mathbf{z}^{(i)}|\mathbf{e}^{(i)})\ln\frac{q(\mathbf{z}^{(i)}|\mathbf{e}^{(i)})}{p(\mathbf{z}^{(i)}|\mathbf{e}^{(i)})}d\mathbf{z}^{(i)}\\
  		=&\mathbb{E}_{q(\mathbf{z}^{(i)}|\mathbf{e}^{(i)})}[\ln\frac{q(\mathbf{z}^{(i)}|\mathbf{e}^{(i)})}{p(\mathbf{z}^{(i)}|\mathbf{e}^{(i)})}]\\
  		=&\mathbb{E}_{q(\mathbf{z}^{(i)}|\mathbf{e}^{(i)})}[\ln q(\mathbf{z}^{(i)}|\mathbf{e}^{(i)})-\ln p(\mathbf{z}^{(i)}|\mathbf{e}^{(i)}) ]\\
  		=&\mathbb{E}_{q(\mathbf{z}^{(i)}|\mathbf{e}^{(i)})}\Big[\ln q(\mathbf{z}^{(i)}|\mathbf{e}^{(i)})-\ln \frac{p(\mathbf{e}^{(i)}|\mathbf{z}^{(i)})p(\mathbf{z}^{(i)})}{p(\mathbf{e}^{(i)})}\Big]\\
  		=&\mathbb{E}_{q(\mathbf{z}^{(i)}|\mathbf{e}^{(i)})}\Big[\ln q(\mathbf{z}^{(i)}|\mathbf{e}^{(i)})-\ln p(\mathbf{e}^{(i)}|\mathbf{z}^{(i)})-\ln p(\mathbf{z}^{(i)})\\
  		&+\ln p(\mathbf{e}^{(i)})\Big]\\
  		=&\mathrm{KL}(q(\mathbf{z}^{(i)}|\mathbf{e}^{(i)})||p(\mathbf{z}^{(i)}))-\mathbb{E}_{q(\mathbf{z}^{(i)}|\mathbf{e}^{(i)})}\Big[\ln p(\mathbf{e}^{(i)}|\mathbf{z}^{(i)})\Big]\\
  		&+\ln p(\mathbf{x}^{(i)}).\\
  	\end{aligned}
  \end{equation}
    As $\ln p(\mathbf{x}^{(i)})$ is a constant due to the deterministic training samples, the optimization object can be rewritten as
    \begin{equation}\label{KL4}
    	\underset{q(\mathbf{z}|\mathbf{e}^{(i)})}{\mathop{\text{minimize}}}\,\quad
    	-\mathbb{E}_{q(\mathbf{z}|\mathbf{e}^{(i)})}[\ln(p(\mathbf{e}^{(i)}|\mathbf{z}))]+ \mathrm{KL}(q(\mathbf{z}|\mathbf{e}^{(i)})||p(\mathbf{z})).
    \end{equation}
\end{appendices}

%
%
%
%

\end{document}